\documentclass[11pt]{article}
\usepackage{amsmath,epsfig,sint}
\font\sixrm=cmr6
\begin{document}

\newcommand{\C}{\rm \kern.25em\vrule height1.4ex
depth-.12ex width.06em\kern-.31em C}

\newcommand{\rme}{{\rm e}}
\newcommand{\rmd}{{\rm d}}
\newcommand{\rmO}{{\rm O}}
\newcommand{\GeV}{{\rm GeV}}
\newcommand{\tr}{{\rm tr}}
\newcommand{\ch}{{\rm ch}}
\newcommand{\sh}{{\rm sh}}
\newcommand{\bz}{\bar{z}}
\newcommand{\On}{\rmO(n)}

\newcommand{\square}{{\sqcup\hskip-2.7mm\sqcap}}

\newcommand{\fY}{{\cal Y}}
\newcommand{\lambdat}{\tilde{\lambda}}

\newcommand{\be}{\begin{equation}}   
\newcommand{\ba}{\begin{eqnarray}}
\newcommand{\ea}{\end{eqnarray}}

\newcommand{\bfk}{{\bf k}}
\newcommand{\bfx}{{\bf x}}
\newcommand{\bfy}{{\bf y}} 
\newcommand{\bfu}{{\bf u}}
\newcommand{\bfv}{{\bf v}}

\newcommand{\wt}{{\widetilde{w}}}
\newcommand{\Mt}{{\widetilde{M}}}
\newcommand{\Wt}{{\widetilde{W}}}

\newcommand{\calK}{{\cal K}}
\newcommand{\hatD}{\hat{D}}
\newcommand{\asinh}{{\rm asinh}}
\newcommand{\gr}{g_{{\hbox{\sixrm R}}}}

\newcounter{subequation}[equation]
\makeatletter

\expandafter\let\expandafter
\reset@font\csname reset@font\endcsname

\def\subeqnarray{\arraycolsep1pt
    \def\@eqnnum\stepcounter##1{\stepcounter{subequation}%
        {\reset@font\rm(\theequation\alph{subequation})}}
\jot5mm     \eqnarray}
\def\endsubeqnarray{\endeqnarray\stepcounter{equation}}

\makeatother

\newcommand{\nspace}{\!\!\!\!\!\!\!\!\!\!}

\newcommand{\msbar}{{\rm \overline{MS\kern-0.14em}\kern0.14em}}

\begin{titlepage}

\begin{flushright}
   MPP-2004-105\\
   September 2004
\end{flushright}

\vskip 0.20 true cm

\begin{center}
{\Large\bf 
Structure functions of the $2d$ O$(n)$ non-linear sigma models}
\end{center}
\vskip 1 true cm
\centerline{\large Janos Balog}
\vskip1ex
\centerline{Research Institute for Particle and Nuclear Physics}
\centerline{1525 Budapest 114, Pf. 49, Hungary}
\vskip 1 true cm
\centerline{\large Peter Weisz}
\vskip1ex
\centerline{Max-Planck-Institut f\"ur Physik}
\centerline{F\"ohringer Ring 6, D-80805 M\"unchen, Germany}
\vskip 1 true cm
\centerline{\bf Abstract}
\vskip 1.0ex
We investigate structure functions in the 2--dimensional 
(asymptotically free) non--linear O$(n)\,\,\sigma$--models 
using the non--perturbative S--matrix bootstrap program.  
In particular the {\it exact small (Bjorken) $x$ behavior is derived}.
Structure functions in the special case of the $n=3$ model are accurately 
computed over the whole $x$ range for $-q^2/M^2<10^5$, and some moments
are compared with results from renormalized perturbation theory.
Some results concerning the structure functions in the $1/n$  
approximation are also presented.

\vfill
\eject

\end{titlepage}


\section{Introduction}

In this paper we study structure functions in 
the asymptotically free $\On$ sigma models in two dimensions.
Due to the integrability of the model one has powerful tools to 
study various non--perturbative properties. In particular one  
can derive the {\it exact small $x$} behavior (for all $q^2$)
and for the case of $n=3$ compute structure functions precisely 
up to very large values of $q^2$. 
Despite the fact that there are no transverse directions,
the structure functions have a rather rich and non--trivial behavior. 
In a previous letter \cite{DIS1} we summarized our results 
and speculated on the possibility of discovering some 
similar structural features in QCD. 

The purpose of this paper is to supply the derivation of the
results presented in \cite{DIS1}.   
This paper is organized as follows. In the next section we give some 
basic definitions of the correlation functions of interest.
In sect.~3 we give the derivation of the (rather universal) exact small
$x$ behavior. Section 4 deals with certain general aspects
concerning the relation of the high $q^2$ behavior of moments of structure
functions to the operator product expansion (OPE). In sect.~5 we 
consider the OPE for the cases of two spin fields and two currents
in the framework of perturbation theory. More detailed results on the
structure functions for the special of $n=3$ are presented in sect.~6.
Finally in sect.~7 we consider computations in the leading order
of the $1/n$ expansion. Many technicalities and some conventions can
be found in the appendices.

\section{$\On$ model and structure functions}

The $\On$ $\sigma$--model in $2d$ (formally described by the Lagrangian
(\ref{onlag})) is perturbatively asymptotically free for $n\ge3$.  
A special property is that these models have
an infinite number of local \cite{Polyakov}
and non--local \cite{Luscher} classical conservation 
laws which survive quantization. 
At the quantum level they imply absence of particle production. 
Assuming the spectrum to consist of one stable 
$\On$--vector multiplet of mass $M$,
the S--matrix has been proposed long ago by the Zamolodchikovs \cite{ZZ}.
Form factors of local operators can be computed 
using general principles \cite{KaWe,Smirnov}. 
The S--matrix bootstrap program for the construction of correlation 
functions involves summing the contributions over all intermediate states
\cite{Karowski}. The possible equivalence of this construction
to the continuum limit of the lattice regularized theory has been
investigated in ref.~\cite{SigmaI}.

\subsection{Current and spin operators, 2--point functions}

The normalization of the conserved $\On$ current operator 
$J^{ab}_\mu(x)$
($a,b=1,\dots n$) is fixed e.g. by the equal time commutation relation 
with the spin field $\Phi^c(y)$:
\be
\left[J^{ab}_0(0,x^1),\Phi^c(0,y^1)\right]=i
\,t^{ab}_{cd}\,\delta(x^1-y^1)\,\Phi^d(0,y^1)\,,
\end{equation}
where the matrices $t^{ab}$ given in (\ref{tdef}) yield the vector 
representation of the $\On$ Lie algebra.  
Its matrix elements are 
\footnote{$\epsilon_{\mu\nu}=-\epsilon_{\nu\mu}\,,\,\,\,\,\epsilon_{01}=1\,$}
\be
\langle0\vert J^{ab}_\mu(0)\vert a_1,\theta_1;\dots;a_r,\theta_r\rangle=
-i\,\epsilon_{\mu\nu}P^\nu_r\,f^{ab}_{a_1\dots a_r}
(\theta_1,\dots,\theta_r)\,.
\label{Currn}
\end{equation}
Here the number of particles, $r$, has to be even and the form factors 
$f^{ab}_{a_1\dots a_r}$ depend on the rapidity
differences only, making Lorentz invariance and current conservation
manifest. The normalization of the $r$--particle states, the
corresponding completeness relations, and other undefined 
kinematics encountered are given in Appendix B.
We define
\be
\sum_{a_1\dots a_r}f^{cd}_{a_1\dots a_r}(\theta_1,\dots,\theta_r)
\,f^{*ef}_{a_1\dots a_r}(\theta_1,\dots,\theta_r)
=\left(\delta^{ce}\delta^{df}-\delta^{cf}\delta^{de}\right)\,C^{(r)}(u)\,,
\label{Ceven}
\end{equation}
where $C^{(r)}(u)$ is a symmetric function of the rapidity
differences.

The normalization of the spin operator $\Phi^a(x)$ is fixed by its
1--particle matrix element:
\be
\langle0\vert\Phi^a(0)\vert b,\theta\rangle=\delta^{ab}\,.
\end{equation}
Its $r$--particle matrix elements ($r$ odd) are defined by
\be
\langle0\vert\Phi^a(0)\vert a_1,\theta_1;\dots;a_r,\theta_r\rangle=
\Lambda_n \,f^a_{a_1\dots a_r}
(\theta_1,\dots,\theta_r)\,,
\end{equation}
where the form factors $f^a_{a_1\dots a_r}$ depend on the rapidity
differences only and the overall factor $\Lambda_n$ is defined
for later convenience. We choose
\be
\Lambda_3=\frac{2}{\sqrt{\pi}},\qquad\qquad
\Lambda_n=1\quad(n>3)\,.
\end{equation}
The analog of (\ref{Ceven}) for odd $r$ is
\be
\sum_{a_1\dots a_r}f^{a}_{a_1\dots a_r}(\theta_1,\dots,\theta_r)
\,f^{*b}_{a_1\dots a_r}(\theta_1,\dots,\theta_r)
=\delta^{ab}\,C^{(r)}(u)\,.
\label{Codd}
\end{equation}

We now make some further definitions: 
\be
I^{(r)}(z)=\frac{1}{(4\pi)^{r-1}}\int{\cal D}u^{(r)}\,
\frac{C^{(r)}(u)}{z+\left[M^{(r)}(u)\right]^2}\,,
\end{equation}
\be
A^{(r)}(z)=-z^2\,\frac{\partial}{\partial z}
I^{(r)}(z)=\frac{1}{(4\pi)^{r-1}}\int{\cal D}u^{(r)}\,
\left(\frac{z}{z+\left[M^{(r)}(u)\right]^2}\right)^2\,C^{(r)}(u)
\label{Adlerfn}
\end{equation}
and for $s=0,1,$
\be
I_s(z)=\sum_{k=0}^\infty I^{(2k+1+s)}(z)\,,
\end{equation}
\be
A_s(z)=\sum_{k=0}^\infty A^{(2k+1+s)}(z)=-z^2\frac{\partial}{\partial z}
I_s(z)\,.
\end{equation}

The invariant functions $I_s$ are related to the 2--point functions
of the current and spin field operators by \cite{JanosMax}
\be
\begin{split}
&\langle0\vert T^*J^{cd}_\mu(x)J^{ef}_\nu(y)\vert0\rangle
=\\
&\quad
\left(\delta^{ce}\delta^{df}-\delta^{cf}\delta^{de}\right)\,
\int\frac{{\rm d}^2p}{(2\pi)^2}\,\rme^{-ip(x-y)}
(p_\mu p_\nu-p^2\eta_{\mu\nu})(-i)I_1(-p^2-i\varepsilon)\,,
\end{split}
\label{MinCurr}
\end{equation}
valid up to contact terms and
\be
\langle0\vert T\Phi^a(x)\Phi^b(y)\vert0\rangle
=\Lambda_n^2\,\delta^{ab}\,
\int\frac{{\rm d}^2p}{(2\pi)^2}\,\rme^{-ip(x-y)}
(-i)I_0(-p^2-i\varepsilon)\,.
\end{equation}
In (\ref{MinCurr}) $T^*$ denotes the covariantized $T$--product.

For $n=3$ we can define $J^a_\mu(x)=\frac{1}{2}\epsilon^{abc}J^{bc}_\mu(x)$
and instead of (\ref{Currn}) we have
\be
\langle0\vert J^{a}_\mu(0)\vert a_1,\theta_1;\dots;a_r,\theta_r\rangle=
-i\,\epsilon_{\mu\alpha}P^\alpha_r\,f^{a}_{a_1\dots a_r}
(\theta_1,\dots,\theta_r)\,.
\label{Curr3}
\end{equation}
In this case instead of (\ref{Ceven}) we can use (\ref{Codd}) also
for $r$ even.

\subsection{Structure functions, moments}

The central object in DIS theory is
\be
W^{ab;cdef}_{\mu\nu}(p,q)=\pi\,\sum_r\,
\langle a,p\vert J^{cd}_\mu(0)\vert r\rangle\,
\langle r\vert J^{ef}_\nu(0)\vert b,p\rangle\,
\delta^{(2)}(p+q-P_r)\,,
\label{W}
\end{equation}
where $q^2<0$. We will use the parameterization
\be
q^2=-4\kappa^2M^2
\label{kappa}
\end{equation}
and the Bjorken variable
\be
x=-\frac{q^2}{2(pq)}\,.
\end{equation} 
Using Lorentz and $\On$ invariance we have
\be
W^{ab;cdef}_{\mu\nu}(p,q)=\left(\eta_{\mu\nu}-\frac{q_\mu q_\nu}{q^2}\right)
\,\sum_{l=0}^2\,R_l^{ab;cdef}\,w_l(q^2,x)\,,
\label{Lor}
\end{equation} 
where the projectors $R_l$ corresponding to the 3 invariant $t$--channel
``isospins" are defined in Appendix A.
Note that in 2 dimensions there is only one independent structure 
function for each isospin channel.

Similarly we define the structure functions corresponding to the spin 
operator through
\be
\Sigma^{ab;cd}(p,q)=-\pi q^2\,\sum_r\,
\langle a,p\vert \Phi^c(0)\vert r\rangle\,
\langle r\vert \Phi^d(0)\vert b,p\rangle\,
\delta^{(2)}(p+q-P_r)\,,
\label{Sigma}
\end{equation}
and
\be
\Sigma^{ab;cd}(p,q)=\Lambda_n^2
\,\sum_{l=0}^2\,P_l^{ab;cd}\,\widetilde w_l(q^2,x)\,,
\end{equation} 
where the $t$--channel projectors $P_l$ for the vector representation
are given in (\ref{pdef0})--(\ref{pdef2}).

Separating the $r$--particle contributions we have
\be
w_l(q^2,x)=\sum_{r {\rm \ odd}}\,w^{(r)}_l(q^2,x)
\qquad {\rm and}\qquad
\widetilde w_l(q^2,x)=\sum_{r {\rm \ even}}\,w^{(r)}_l(q^2,x)
\end{equation} 
with
\be
w^{(r)}(q^2,x)=\frac{-\pi q^2}{(4\pi)^r}\,\int_{-\infty}^\infty
{\rm d}\Lambda\int{\cal D}u^{(r)}\,\delta^{(2)}(p+q-P_r)\,
J^{(r)}_l(\theta)\,.
\label{wr1}
\end{equation} 
Here $p=(M,0)$ and for $r$ odd
\be
J^{(r)}_l(\theta)=\frac{1}{\pi_l\hat r_l}\,
\sum_{\stackrel{abcdef}{a_1\dots a_r}}\,R_l^{ab;cdef}\,
f^{cd}_{aa_1\dots a_r}(i\pi,\theta_1,\dots,\theta_r)
\,f^{*ef}_{ba_1\dots a_r}(i\pi,\theta_1,\dots,\theta_r)\,,
\label{Jodd}
\end{equation}
while for $r$ even
\be
J^{(r)}_l(\theta)=\frac{1}{\pi_l}\,
\sum_{\stackrel{abcd}{a_1\dots a_r}}\,P_l^{ab;cd}\,
f^{c}_{aa_1\dots a_r}(i\pi,\theta_1,\dots,\theta_r)
\,f^{*d}_{ba_1\dots a_r}(i\pi,\theta_1,\dots,\theta_r)\,.
\label{Jeven}
\end{equation}

By doing the $\Lambda$--integration we can further simplify (\ref{wr1}):
\be
w^{(r)}_l(q^2,x)=\frac{2\kappa^2}{(4\pi)^{r-1}}\int{\cal D}u^{(r)}\,
\delta\left[\mu_r^2-1-\frac{4\kappa^2}{x}+4\kappa^2\right]
J^{(r)}_l(\bar{\beta}_1,\dots,\bar{\beta}_r)\,,
\label{wr2}
\end{equation} 
where
\ba
\bar{\beta}_j&=&\beta_j+b+2v^{(r)}_-\,,
\\ 
b&=&\ln\left\{\frac{1}{2}+\frac{1}{2}\sqrt{1+\frac{x^2}{\kappa^2}}+
\frac{x}{4\kappa^2}\right\}-\ln\left\{1-x+\frac{x}{4\kappa^2}\right\}\,.
\label{defb}
\ea

We define the structure function moments by
\be
M_{l;N}(q^2)=\int_0^1{\rm d}x\,x^{N-1}\,w_l(q^2,x)
\quad{\rm and}\quad
\widetilde M_{l;N}(q^2)=\int_0^1{\rm d}x\,x^{N-1}\,
\widetilde w_l(q^2,x)
\label{moments}
\end{equation} 
and similarly for fixed particle number
\be
M^{(r)}_{l;N}(q^2)=\int_0^1{\rm d}x\,x^{N-1}\,w^{(r)}_l(q^2,x)\,.
\label{fixmoments}
\end{equation} 
Obviously
\be
M_{l;N}(q^2)=\sum_{r{\rm \ odd}}M^{(r)}_{l;N}(q^2)
\qquad\quad{\rm and}\quad\qquad
\widetilde M_{l;N}(q^2)=\sum_{r{\rm \ even}}M^{(r)}_{l;N}(q^2)\,.
\end{equation} 

The $r$--particle moments can also be calculated directly from
(\ref{wr2}):
\ba
M^{(r)}_{l;N}(q^2)&=&\frac{1}{2(4\pi)^{r-1}}\int{\cal D}u^{(r)}\,\left[x^{N+1}
J^{(r)}_l(\bar{\beta}_1,\dots,\bar{\beta}_r)\right]_{x=\bar{x}}\,,
\label{Mr}
\\
\bar{x}&=&\frac{4\kappa^2}{4\kappa^2+\mu_r^2-1}\,.
\ea

For $n=3$ (\ref{W}) can be written as
\be
W^{ab;cd}_{\mu\nu}(p,q)=\pi\,\sum_r\,
\langle a,p\vert J^{c}_\mu(0)\vert r\rangle\,
\langle r\vert J^{d}_\nu(0)\vert b,p\rangle\,
\delta^{(2)}(p+q-P_r)
\label{WO3}
\end{equation}
and (\ref{Lor}) becomes
\be
W^{ab;cd}_{\mu\nu}(p,q)=\left(\eta_{\mu\nu}-\frac{q_\mu q_\nu}{q^2}\right)
\,\sum_{l=0}^2\,P_l^{ab;cd}\,w_l(q^2,x)\,.
\end{equation} 
In this case (\ref{Jeven}) is valid for odd as well as even $r$ values
(with $\pi_l=2l+1$).

The 2--particle form factor can be written
\be
f^{cd}_{ab}(\theta_1,\theta_2)=\phi(\theta_1-\theta_2)\,
\left(\delta^{ac}\delta^{bd}-\delta^{ad}\delta^{bc}\right)\,,
\label{phi}
\end{equation} 
with
\be
\phi(\theta)=-\tanh\frac{\theta}{2}\,\,
\exp\left\{-2\int_0^{\infty}{\rmd t \over t}
{\left[1-\rme^{-\frac{2t}{n-2}}\over 1+\rme^t\right]}
{\sin^2([i\pi-\theta]t/2\pi)\over\sinh t}\right\}\,.
\label{j2partffex}
\end{equation}
The 1--particle contribution to the structure functions
is then given by
\be
w^{(1)}_l(q^2,x)=m_l\,\delta(x-1)\,\vert\phi(i\pi-\alpha)\vert^2\,, 
\end{equation} 
where 
\be
\sinh\frac{\alpha}{2}=\kappa\,,\quad\qquad{\rm and}\qquad\quad
m_0=1,\quad m_1=-m_2=\frac{1}{2}\,.
\label{ml}
\end{equation}

\section{$2d$ structure functions at small $x$}

In this section we derive a general formula describing the asymptotic 
behavior of the $\On$ model structure functions at small $x$ values.
The derivation is based on general properties of the form factors and
the scattering matrix elements and therefore the behavior we find here is
expected to hold in other $2d$ integrable models as well.

For small $x\to0$ the variable $b$ in (\ref{defb}) behaves as
\be
b=x+\rmO(x^2)\,,
\end{equation} 
and if we do the $u_{r-1}$-integration in (\ref{wr2}) with the help of
the delta function we get
\be
u_{r-1}=\ln\left(\frac{4\kappa^2}{x}\right)-2v^{(r-1)}_++
\rmO(x)\,,
\end{equation} 
and further
\be
2v^{(r)}_-=\frac{x\mu_{r-1}^2}{4\kappa^2}+\rmO(x^2),\qquad
\frac{\partial\mu_r^2}{\partial u_{r-1}}=\frac{4\kappa^2}{x}+\rmO(1)\,.
\end{equation} 
Now putting all the above together we have
\be
w^{(r)}_l(q^2,x)\cong \frac{x}{2(4\pi)^{r-1}}\int{\cal D}u^{(r-1)}\,
J^{(r)}_l(-\varepsilon,
-\beta-\varepsilon +\widetilde \beta_{r-1},\dots,
-\beta-\varepsilon +\widetilde \beta_1)\,,
\label{smx1}
\end{equation} 
where
\be
\varepsilon=-\left(1+\frac{\mu_{r-1}^2}{4\kappa^2}\right)x+\rmO(x^2)
\qquad\quad{\rm and}\qquad\quad \beta=\ln x+\rmO(1)
\end{equation} 
and $\widetilde \beta_i$ are the variables for $r-1$ particles.

Eqns.~(\ref{Jodd}) and (\ref{Jeven}) are both of the form
\be
J^{(r)}_l(\theta)=
\sum_{\stackrel{abAB}{a_1\dots a_r}}\,C^{ab;AB}\,
f^A_{aa_1\dots a_r}(i\pi,\theta_1,\dots,\theta_r)
\,f^B_{ba_1\dots a_r}(i\pi,\theta_1,\dots,\theta_r)\,,
\label{Jgen}
\end{equation}
where for the current case
\be
C^{ab;AB}=\frac{1}{\pi_l\hat r_l}\,R_l^{ab;cdef}\qquad\quad
{\rm with}\quad\qquad
A\sim cd\,,\qquad B\sim ef\,,
\end{equation} 
and for the spin case
\be
C^{ab;AB}=\frac{1}{\pi_l}\,P_l^{ab;cd}\qquad\quad
{\rm with}\quad\qquad
A\sim c\,,\qquad B\sim d\,.
\end{equation} 
With this notation we can write $J^{(r)}_l$ in (\ref{smx1}) as
\be
\begin{split}
J^{(r)}_l=
\sum_{\stackrel{abAB}{a_1\dots a_r}}\,C^{ab;AB}\,
f^A_{aa_r\dots a_1}(i\pi+\varepsilon &+\beta,\beta,
\widetilde\beta_{r-1},\dots,\widetilde\beta_1)\cdot\\
&\cdot f^{*B}_{ba_r\dots a_1}(i\pi+\varepsilon+\beta,\beta,
\widetilde\beta_{r-1},\dots,\widetilde\beta_1)\,.
\end{split}
\label{smx2}
\end{equation}
The crucial point now is that
since $\varepsilon$ is small and $\beta$ is large we can here use
(\ref{asyRES2}) which follows from general principles encoded
in the Smirnov axioms. In leading order we get
\be
\begin{split}
J^{(r)}_l\cong \frac{(4\pi)^2}{(n-2)^2\varepsilon^2\beta^2}\,
\sum_{\stackrel{abAB}{a_1\dots a_r}}\,C^{ab;AB}\,
&t^{aa_r}_{AA^\prime}\,t^{ba_r}_{BB^\prime}\,
f^{A^\prime}_{a_{r-1}\dots a_1}(
\widetilde\beta_{r-1},\dots,\widetilde\beta_1)\cdot\\
&\cdot f^{*B^\prime}_{a_{r-1}\dots a_1}(
\widetilde\beta_{r-1},\dots,\widetilde\beta_1)\,,
\end{split}
\label{smx3}
\end{equation}
which can be further simplified with the help of (\ref{Vdef}),
(\ref{Tdef}), (\ref{Ceven}) and (\ref{Codd}) leading to
\be
J^{(r)}_l\cong \frac{(4\pi)^2}{(n-2)^2\varepsilon^2\beta^2}\,
G_l\,C^{(r-1)}(u)\,,
\end{equation}
where the constants $G_l$ are equal to $V_l$ and $T_l$ for the spin 
and current cases respectively given in (\ref{vl}) 
and (\ref{tl}), and further
\be
w^{(r)}_l(q^2,x)\cong\frac{1}{x\ln^2x}\,
\frac{8\kappa^4G_l}{(4\pi)^{r-3}(n-2)^2}\,
\int{\cal D}u^{(r-1)}\frac{C^{(r-1)}(u)}
{\left(4\kappa^2+\mu_{r-1}^2\right)^2}\,,
\end{equation} 
which can also be written as
\be
w^{(r)}_l(q^2,x)\cong\frac{1}{x\ln^2x}\,\frac{2\pi G_l}{(n-2)^2}\,
A^{(r-1)}(-q^2)\,,
\end{equation} 
where the Adler functions $A^{(r)}$ were defined in (\ref{Adlerfn}).

The final results for the complete structure functions are
\be
w_l(q^2,x)\cong\frac{1}{x\ln^2x}\,\frac{2\pi T_l}{(n-2)^2}\,
A_1(-q^2)
\label{smxT}
\end{equation} 
and
\be
\widetilde w_l(q^2,x)\cong\frac{1}{x\ln^2x}\,\frac{2\pi V_l}{(n-2)^2}\,
A_0(-q^2)\,,
\label{smxV}
\end{equation} 
Note that the structure of the asymptotic small $x$ behavior, 
factorizing a part characteristic
to the target and a part described by the vacuum 2--point function,
is rather universal being independent of the operator, 
independent of $n$, and independent of the isospin channel.  

This completes the derivation of the exact small $x$ asymptotics
first announced in ref.~\cite{DIS1}. The question of
possible lessons that can be learned for QCD was addressed in the
latter reference and will not be repeated here.

\section{The operator product expansion}

In the $\On\,\,\sigma$--models there does not seem to be a simple parton
picture. This is even so for the case $n=3$ where the model is equivalent
to the $\C{\rm P}^1$ model. For although this model is formulated in 
terms of a complex doublet of fields which are analogous to  
quarks in that they are confined, it seems that they do not play a r\^{o}le 
more similar to partons than the elementary bare spin fields in the 
original formulation
\footnote{Perhaps the peculiar threshold behavior discussed in sect.~6 
is explained by the fact that (as opposed to QCD) with some probability
the $\On$ particle can consist of a single point--like
parton that carries the same quantum numbers.}. 
The question is related to that of understanding  
what are (if any) the ``ultra--particles" in the sense 
of Buchholz and Verch \cite{BuVe}, 
or to the associated question as to whether the 
$\sigma$--models have an underlying conformal field theory. 

Although an intuitive parton description with suggestive DGLAP equations
\be
q^2\frac{\partial}{\partial q^2}w_l(q^2,x)
=\int_x^1\frac{\rmd y}{y} p_l(x/y,q^2)w_l(q^2,y)\,,
\label{ateqtn}
\end{equation}
(where $p_l(z,q^2)$ would be the corresponding splitting functions) 
is still missing in these models, we still have the machinery of the 
operator product expansion (OPE) to give us information on 
the evolution of the moments (\ref{moments}) at large $-q^2$.

The OPE in the sigma model is surprisingly involved and hence we have 
decided to present the material as follows. In the next subsection
we first summarize the results; readers who would prefer to skip
the derivations can then jump to sect.~6. The general structure
of the product of two local operators (in this case the spins and currents)
is described in the remaining part of this section. 
Our analysis extends that initiated e.g. in refs.~\cite{Luscher}, 
\cite{CMP}. 
So far too little is rigorously known about the detailed structure
of the OPE from the general principles of the bootstrap approach
to obtain the explicit results below. The extra required information 
is however supplied in the framework of renormalized perturbation 
theory which is presented in sect.~5. Some comparisons of the moments
with those from the bootstrap approach at high $-q^2$ are 
presented in subsect.~6.4.

\subsection{Summary of results on the moments}

For the current ($N$ even) moments in the isospin 0 channel we have
\be
M_{0;N}(q^2)=W_{0;N}\,\frac{n-2}{2(n-1)}
\,\left\{1+\frac{1}{n-2}\lambda(q^2)+\rmO\left(\lambda^2\right)\right\}\,,\,\,\,
N\ge2\,,
\label{curr0J}
\end{equation}
where $\lambda(q^2)$ is an effective running coupling function defined 
through
\be
\frac{1}{\lambda(q^2)}+\frac{1}{n-2}\ln\lambda(q^2)
=\ln \frac{\sqrt{\vert q^2\vert}}{\Lambda_\msbar}\,,
\label{deflambda}
\end{equation}
and the $W_{0;N}$ are
renormalization group invariant, non--perturbative constants,
corresponding to the matrix elements of spin $N$ operators. 
In the $N=2$ case this is the energy--momentum tensor 
operator $T_{\mu\nu}$ for which we know the constant explicitly
\be
\langle a,p|T_{\mu\nu}(0)|b,p\rangle=
W_{0;2}p_\mu p_\nu\delta^{ab}\,,\,\,\,\,\,\,\,W_{0;2}=2\,. 
\end{equation}
In particular the ``momentum sum rule" follows:
\be
M_{0;2}(-\infty)=\frac{n-2}{n-1}\,.
\label{m02infty}
\end{equation}
Note that all the isospin 0 moments tend to constants as 
$-q^2\to\infty$.
As a consequence these current structure functions in the $\On$ models 
obey Bjorken scaling. Computations in the $n=3$ model, (see sect.~6
and in particular Fig.~3), indicate that the resulting limiting
scaling functions are non--trivial.
This is a special property of these models and 
we conjecture that this is due to the existence of
an infinite set of local conserved quantities \cite{Polyakov}. 

In the isospin $l=1$ channel for odd moments $N\ge3$ 
we can only say that
\be
M_{1;N}(q^2)=W_{1;N}\,\lambda(q^2)^{\frac{1}{n-2}}+\dots\,,\,\,N\ge 3\,,
\label{curr1J}
\end{equation}
but in the special case $N=1$ we have
\be
M_{1;1}(q^2)=\frac{1}{2}\,
\,\left\{1-\frac{1}{n-2}\lambda(q^2)+\rmO\left(\lambda^2\right)\right\}\,,
\label{curr11}
\end{equation}
where the constant is known through the current normalization
\be
\langle a,p|J_\mu^{cd}(0)|b,p\rangle=-4ip_\mu P_1^{ab;cd}\,.
\label{currentnorm}
\end{equation}
From this follows the analogy to the Adler sum rule in QCD:
\be
M_{1;1}(-\infty)=\frac12\,.
\label{m11infty}
\end{equation}

For the spin field isospin 0 moments we have 
\ba
\Mt_{0;N}(q^2)&=&
\frac{W_{0;N}\pi^2nC_n}{(n-2)^2}\lambda(q^2)^{\frac{n-3}{n-2}}
\left\{1+\rmO(\lambda)\right\}\,,\,\,\,\,n\ge4\,,
\label{Mt0N}
\\
\Mt_{0;N}(q^2)&=&
\frac{W_{0;N}}{4}\left\{1+\lambda(q^2)+\rmO(\lambda^2)\right\}
\,,\,\,\,\,\,\,\,\,\,\,\,\,\,\,\,\,\,n=3\,,
\label{spin0J}
\ea
where the non--perturbative constants $W_{0;N}$ are the same as for the 
current, and where $C_n$ is the non--perturbative constant appearing 
in the short distance expansion 
\be
\langle0|\Phi^a(y)\Phi^b(0)|0\rangle\sim 
C_n\delta^{ab}\left(-\ln M|y|\right)^{\frac{n-1}{n-2}}\,.
\label{Cn}
\end{equation}
So far the value of $C_n$ is not known for general $n$;
for the case $n=3$ a (well tested) conjecture based on scaling 
\cite{JanosMax} gives
\be
C_3=\frac{1}{3\pi^3}\,,
\end{equation}
and we know for $n=\infty$:
\be
C_\infty=\frac{1}{2\pi}\,.
\end{equation}
We see that only for the case $n=3$ do the moments of the 
field $l=0$ structure function have the same leading asymptotic 
behavior as those of the current.

For the isospin $l=1$ field (odd) moments we find to leading order PT  
\ba
\Mt_{1;1}(q^2)&=&\Mt_{0;2}(q^2)\,,
\label{Mt11}
\\
\Mt_{1;N}(q^2)&=&\Wt_{1;N}\lambda(q^2)^{\frac{2n-5}{n-2}}
\left\{1+\rmO\left(\lambda^{\frac{1}{n-2}}\right)\right\}\,,\,\,N\ge3\,,
\label{spin1J}
\ea
where there is in general no obvious relation between the 
$\Wt_{1;N}$ and the constants occurring in (\ref{curr1J}), except 
for $n=3$ where they are equal ($\Wt_{1;N}=W_{1;N}\,,\,\,n=3)$.

For isospin $l=2$ moments we obtain (for all $n\ge3$):
\be
\Mt_{2;N}(q^2)=
\Wt_{2;N}\lambda(q^2)^2
\left\{1+\rmO\left(\lambda^{\frac{1}{n-2}}\right)\right\}\,.
\end{equation}

Finally using the exact ratio of the mass to the $\Lambda$--parameter
\be
\frac{M}{\Lambda_\msbar}=\frac{(8/\rme)^{1/(n-2)}}{\Gamma[1+1/(n-2)]}\,,
\label{Lambdaoverm}
\end{equation}
obtained by Hasenfratz, Maggiore and Niedermayer \cite{HMN},
the perturbative results can be plotted as functions of $-q^2/M^2$.

\subsection{Dispersion relations}

For the discussion of the OPE it is convenient to work in
the Euclidean formalism. For local operators ${\cal A}$ we have
\be
{\cal A}(x^0,x^1)=
\rme^{i(Hx^0-Px^1)}\,{\cal A}(0,0)\,\rme^{-i(Hx^0-Px^1)}\,,
\end{equation}
where $H,P$ are the time and spatial translation operators.
We can similarly define Euclidean translation by
\be
{\cal A}^{\rm E}(y_1,y_2)=
\rme^{Hy_2-iPy_1}\,{\cal A}(0,0)\,\rme^{-Hy_2+iPy_1}\,,
\end{equation}
which is formally ${\cal A}(-iy_2,y_1)$. For Euclidean vectors (and 
similarly for tensors) we define
\be
V_2=-iV_0.
\end{equation}

The Euclidean time ordering is defined as
\be
\begin{split}
T_{\rm E}\,\left({\cal A}^{\rm E}(y_1,y_2){\cal B}^{\rm E}(z_1,z_2)\right)
&=\Theta(y_2-z_2)\,{\cal A}^{\rm E}(y_1,y_2){\cal B}^{\rm E}(z_1,z_2)\\
&+\Theta(z_2-y_2)\,{\cal B}^{\rm E}(z_1,z_2){\cal A}^{\rm E}(y_1,y_2)\,,
\end{split}
\end{equation}
and the connected part of the product of two operators is
\be
\left({\cal O}_1{\cal O}_2\right)_c=
{\cal O}_1{\cal O}_2-\langle0\vert
{\cal O}_1{\cal O}_2\vert0\rangle-
{\cal O}_1\vert0\rangle\langle0\vert{\cal O}_2\,.
\end{equation}

We now define Euclidean functions for the currents: 
\be
\begin{split}
&\frac{1}{2}\,\int\rmd y_1\,\rmd y_2\,\rme^{i(Q_1y_1+Q_2y_2)}
\langle a,0\vert\,T_{\rm E}^*\left(J^{cd{\rm E}}_\mu(y_1,y_2)\,
J^{ef}_\nu(0,0)\right)_c\vert b,0\rangle\\
&=\left(Q_\mu Q_\nu-Q^2\delta_{\mu\nu}\right)\,\sum_{l=0}^2\,
R_l^{ab;cdef}\,\tau_l(Q^2,Q_2)\,,
\end{split}
\end{equation}
where $T^*_{\rm E}$ stands for covariantized Euclidean time ordering,
i.e. some non--covariant terms proportional to delta functions
of the Euclidean time difference (and derivatives of the delta function)
are dropped. 

Similarly for the spin field:
\be
\begin{split}
&\frac{1}{2}\,\int\rmd y_1\,\rmd y_2\,\rme^{i(Q_1y_1+Q_2y_2)}
\langle a,0\vert\,T_{\rm E}\left(\Phi^{c{\rm E}}(y_1,y_2)\,
\Phi^d(0,0)\right)_c\vert b,0\rangle\\
&=\Lambda_n^2\,\sum_{l=0}^2\,
P_l^{ab;cd}\,\tilde\tau_l(Q^2,Q_2)\,.
\end{split}
\end{equation}

The $\tau_l$ and $\tilde{\tau}_l$, as functions
of $Q_2$ at fixed real $Q^2$ are real analytic
\be
\tau_l(Q^2,Q_2)^*=\tau_l(Q^2,-Q_2^*)\,,\,\,\,\,\,\,\,\,
\tilde{\tau}_l(Q^2,Q_2)^*=\tilde{\tau}_l(Q^2,-Q_2^*)\,,
\end{equation}
and obey the crossing properties
\be
\tau_l(Q^2,Q_2)=(-1)^l\tau_l(Q^2,-Q_2)\,,\,\,\,\,\,\,\,\,
\tilde{\tau}_l(Q^2,Q_2)=(-1)^l\tilde{\tau}_l(Q^2,-Q_2)\,.
\end{equation}
Further they have cuts along parts of the imaginary axis
(with poles at $Q_2=\pm iQ^2/2M$), 
and the discontinuities across the cuts are simply
related to the structure functions:
\ba
w_l(-Q^2,x)&=& \frac{Q^2}{\pi}\,{\rm Im}\,
\tau_l\left(Q^2,\varepsilon-i\frac{Q^2}{2Mx}\right)\,,\\
\wt_l(-Q^2,x)&=& \frac{Q^2}{\pi}\,{\rm Im}\,
\tilde{\tau}_l\left(Q^2,\varepsilon-i\frac{Q^2}{2Mx}\right)\,.
\ea

Concerning the general singularity structure in the complex $Q_2$ plane 
away from the imaginary axis,
little more is rigorously known except that the current function 
$\tau_1$ has poles on the real $Q_2$ axis originating from the 
contribution from 1--particle intermediate states. 
The contribution from 1--particle states is easily computed:
\be
\begin{split}
\tau^{{\rm 1-part}}_l(Q^2,Q_2)&=
-\frac{m_l}{2M\cosh k}\phi(i\pi-k)\phi(i\pi+k)\times\\
&\left\{\frac{1}{M(\cosh k-1)-iQ_2}
+\frac{(-1)^l}{M(\cosh k-1)+iQ_2}\right\},
\end{split}
\end{equation}
where $Q_1=M\sinh k$, $\phi(\theta)$ is the form factor function 
(\ref{j2partffex}) and the constants $m_l$ are given in (\ref{ml}).
Since for small $k$
\be
\phi(i\pi+k)\cong -\frac{2}{k}\,,
\end{equation}
the 1--particle contribution, for fixed $Q^2$ as function of $Q_2$,
has poles at $Q_2=\pm \sqrt{Q^2}$ with residue 
$-\frac{iM}{Q^2}\delta_{l1}$.

Assuming that no other singularities are generated by the higher
intermediate states away from the imaginary $Q_2$ axis, from the usual 
Cauchy integral we conclude that for a circular contour centered at the 
origin with radius $\sqrt{Q^2}<R<Q^2/(2M)$
\be
\frac{1}{2\pi i}\oint\frac{\rmd\zeta\,\tau_l\left(Q^2,\zeta\right)}
{\zeta^{N+1}}=i^N\tau_{l;N}\left(Q^2\right)-i\frac{2M}{Q^2}\delta_{l1}\frac{1}
{(\sqrt{Q^2})^{N+1}}\,,
\end{equation}
where
$\tau_{l;N}(Q^2)$ are the coefficients of the Taylor expansion 
\be
\tau_l(Q^2,Q_2)=\sum_{N=0}^\infty\tau_{l;N}(Q^2)\,(iQ_2)^N\,.   
\end{equation}
Now the structure function moments can be computed in the usual way by
calculating the Cauchy integral along the deformed contour
around the cuts. In this way we obtain expressions for the moments:
\be
M_{l;N}\left(-Q^2\right)=M\left(\frac{Q^2}{2M}\right)^{N+1}\left\{
\tau_{l;N}\left(Q^2\right)+
\frac{2M}{Q^2}\delta_{l1}\frac{1}{(i\sqrt{Q^2})^{N+1}}\right\}\,.
\label{w1Ncauchy}
\end{equation}
The spin function $\tilde{\tau}$ has no 1--particle 
contribution, and so assuming no further singularities apart 
from the cuts we obtain for the moments
\be
\widetilde M_{l;N}(-Q^2)=M\left(\frac{Q^2}{2M}\right)^{N+1}\, 
\tilde{\tau}_{l:N}(Q^2)\,,
\end{equation}
where 
\be
\tilde{\tau}_l(Q^2,Q_2)=\sum_{N=0}^\infty\tilde{\tau}_{l;N}(Q^2)\,(iQ_2)^N\,.   
\end{equation}

Note in the equations above for $l=0,2$ $N$ is even, positive and for 
$l=1$ $N$ is odd.
It remains to extract information on the Taylor coefficients 
$\tau_{l;N},\tilde{\tau}_{l;N}$ from the operator product expansions.

\subsection{Operator product expansion for the spin field}

Starting with the spin field, the
connected part of the time ordered product can be expanded as:
\be
\begin{split}
T_{\rm E}&\left(\Phi^{c{\rm E}}(y_1,y_2)\,\Phi^d(0,0)\right)_c=\\
&\qquad
\sum_{l,\omega}{\cal A}^{(l)cd}_\omega\,\gamma^{(l)}_\omega(y^2)
+\sum_{J=1}^\infty\,\sum_{l,\omega}
\,\gamma^{(J,l)}_\omega(y^2)\left\{ {\cal B}^{(J,l)cd}_\omega\,y_+^J
+\overline{{\cal B}}^{(J,l)cd}_\omega\,y_-^J\right\}\,,
\end{split}
\label{OPE}
\end{equation}
where $y_\pm=\mp y_1-iy_2$.
Employing a basis of hermitian operators 
\be
{{\cal A}_\omega^{(l)cd}}^\dagger={\cal A}^{(l)cd}_\omega\,,\qquad
{{\cal B}_\omega^{(J,l)cd}}^\dagger={\cal B}^{(J,l)cd}_\omega
\end{equation}
and using Poincar\'e symmetry, parity and CPT invariance we have
\be
\overline{{\cal B}}^{(J,l)cd}_\omega=V\,{\cal B}^{(J,l)cd}_\omega\,V\,,
\qquad\qquad\quad
{\cal A}^{(l)cd}_\omega=V\,{\cal A}^{(l)cd}_\omega\,V\,,
\end{equation}
where $V$ is the parity operator and
\be
{\gamma_\omega^{(l)}}^*(y^2)=\gamma^{(l)}_\omega(y^2),\qquad\qquad\quad
{\gamma_\omega^{(J,l)}}^*(y^2)=\gamma^{(J,l)}_\omega(y^2).
\end{equation}
Further we define the matrix elements $B^{(J,l)}_{\omega}$ as in
appendix \ref{appE}
\be
\langle a,\theta\vert\,{\cal B}^{(J,l)cd}_\omega\vert b,\theta\rangle=
\left(-i\frac{M}{2}\rme^\theta\right)^J\,
P_l^{ab;cd}\,B^{(J,l)}_{\omega}
\end{equation}
and we find
\be
{B^{(J,l)}_{\omega}}^*=B^{(J,l)}_{\omega}=
(-1)^{J+l}\,B^{(J,l)}_{\omega}\,.
\end{equation}

The \lq\lq twist" of the operator is defined as
\be
t^{(J,l)}_\omega={\rm dim}\,\left({\cal B}^{(J,l)cd}_\omega\right)-J
\end{equation}
and the minimal possible twist value is zero. The contribution of
these operators dominate for large momenta and we have
\be
\widetilde 
M_{l;N}(-Q^2)\approx\frac{1}{4\Lambda_n^2}\,\hat\eta^{(N,l)}(Q^2)\,,
\end{equation}
where
\ba
\hat\eta^{(J,l)}(Q^2)&=&(Q^2)^{J+1}\,\left(\frac{\rmd}{\rmd Q^2}\right)^J\,
\int \,\rmd^2y\,\rme^{iQy}\,\eta^{(J,l)}(y^2)\,,
\\
\eta^{(J,l)}(y^2)&=&\sum_\omega\,B^{(J,l)}_{\omega}\,
\gamma^{(J,l)}_\omega(y^2)\,.
\label{eta}
\ea
 
\subsection{Operator product expansion for the current}

Using hermicity, Poincar\'e, $\On$, parity and CPT symmetries and
current conservation we can write
\be
\begin{split}
T_{\rm E}\,\left(J^{cd{\rm E}}_\mu(y)\,J^{ef}_\nu(0)\right)
&=\sum_{\stackrel{l=0,2}{\omega}}R^{ab;cdef}_l\,
H^{(l)}_{\mu\nu;\omega}(y)\,{\cal A}^{(l)ab}_\omega+\\
&\sum_{\stackrel{J\equiv l}{\omega}}\,R^{ab;cdef}_l\,\left\{
H^{(J,l)}_{\mu\nu;\omega}(y)\,{\cal B}^{(J,l)ab}_\omega+
\overline{H}^{(J,l)}_{\mu\nu;\omega}(y)\,
\overline{{\cal B}}^{(J,l)ab}_\omega\right\}\,\dots
\end{split}
\end{equation}
where the dots indicate that we have omitted total derivative
operators since they would not contribute to the diagonal expectation 
values. Otherwise the set of operators appearing here is as in
(\ref{OPE}) and the 
coefficient functions $H^{(l)}_{\mu\nu;\omega}$ take the form
\be
\begin{align}
H^{(l)}_{++;\omega}(y)&=-\frac{y_-}{y_+}
\fY(\fY+1)V^{(l)}_\omega\left(y^2\right),\\
H^{(l)}_{+-;\omega}(y)&=H^{(l)}_{-+;\omega}(y)=
(\fY+1)^2V^{(l)}_\omega\left(y^2\right),\\
H^{(l)}_{--;\omega}(y)&=-\frac{y_+}{y_-}
\fY(\fY+1)V^{(l)}_\omega\left(y^2\right),
\end{align}
\end{equation}
where $V^{(l)}_\omega\left(y^2\right)$ ($l=0,2$) are real functions unique
up to $\frac{\rm const.}{y^2}$, and 
\be
\fY=y^2\frac{\rmd}{\rmd y^2}\,.
\end{equation}
Similarly
\be
\begin{align}
H^{(J,l)}_{++;\omega}(y)&=\frac{c_\omega}{y_+}\delta_{J1}
-y_-y_+^{J-1}(\fY+J)(\fY+J+1)V^{(J,l)}_\omega\left(y^2\right),\\
H^{(J,l)}_{+-;\omega}(y)&=H^{(J,l)}_{-+;\omega}(y)=
y_+^J(\fY+1)(\fY+J+1)V^{(J,l)}_\omega\left(y^2\right),\\
H^{(J,l)}_{--;\omega}(y)&=-\frac{y_+^{J+1}}{y_-}
\fY(\fY+1)V^{(J,l)}_\omega\left(y^2\right),
\end{align}
\end{equation}
where $V^{(J,l)}_\omega\left(y^2\right)$ ($J\equiv l$) are real 
functions unique up to $\frac{\rm const.}{y^2}\delta_{J1}$ and 
$c_\omega$ are real constants.
Finally
\be
\overline{H}^{(J,l)}_{\mu\nu;\omega}(y_+,y_-)=
H^{(J,l)}_{\bar\mu \bar\nu;\omega}(y_-,y_+),
\end{equation}
where $\bar\mu=-\mu$ for the light-cone index $\mu=\pm$.

As a consequence of the asymptotic freedom of the $\On$ model for
small $y^2$
\be
\begin{align}
V^{(l)}_\omega\left(y^2\right)&\sim \vert y\vert^{t^{(l)}_\omega-2},\\
V^{(J,l)}_\omega\left(y^2\right)&\sim \vert y\vert^{t^{(J,l)}_\omega-2},
\end{align}
\end{equation}
where $t^{(l)}_\omega, t^{(J,l)}_\omega$ are the twist of the
corresponding operators.

\subsubsection{Fourier transformation}

Introducing
\be
\begin{align}
X^{(l)}_\omega(y)&=y^2V^{(l)}_\omega(y^2),\\
X^{(J,l)}_\omega(y)&=y^2y_+^JV^{(J,l)}_\omega(y^2),\\
\overline{X}^{(J,l)}_\omega(y)&=y^2y_-^JV^{(J,l)}_\omega(y^2)
\end{align}
\end{equation}
the current operator product in Fourier space can be written
\ba
&&\int\rmd^2Q\,\rme^{iQy}\langle a,0\vert
T_{\rm E}\,\left(J^{cd{\rm E}}_\mu(y)\,J^{ef}_\nu(0)\right)
\vert b,0\rangle
=-\sum_{\stackrel{l=0,2}{\omega}}
R^{ab;cdef}_l E_{\mu\nu}(Q)\tilde 
X_\omega^{(l)}(Q)A^{(l)}_\omega\label{FT}
\nonumber\\
&&-\sum_{\stackrel{J\equiv l}{\omega}}R_l^{ab;cdef}\left\{
E_{\mu\nu}(Q)\left[\tilde X^{(J,l)}_\omega(Q)+ 
\tilde{\overline{X}}^{(J,l)}_\omega(Q)\right]
+i\pi c_\omega\delta_{J1}K_{\mu\nu}(Q)\right\}
\left(\frac{-iM}{2}\right)^J B^{(J,l)}_\omega
\nonumber\\
&&
\ea
where the reduced matrix elements $A^{(l)}_\omega, B^{(J,l)}_\omega$ 
are real and $E_{\mu\nu}(Q)$ is the transversal
tensor
\be
E_{\mu\nu}(Q)=Q_\mu Q_\nu-Q^2\delta_{\mu\nu}\,.
\end{equation}
The complete expression (\ref{FT}), although conserved in 
coordinate space, is not transversal because of the anomalous
terms proportional to the constants $c_\omega$. These
are multiplied by the tensor $K_{\mu\nu}(Q)$ with components
\be
K_{++}=\frac{1}{Q_-},\qquad\qquad
K_{--}=\frac{1}{Q_+},\qquad\qquad
K_{+-}=K_{-+}=0.
\end{equation}
It is not quite trivial to see, but easy to check that
\be
K_{\mu\nu}(Q)=-E_{\mu\nu}(Q)\frac{4iQ_2}{Q_1^2Q^2}+\frac{\mu+\nu}{Q_1}
-\mu\nu\frac{iQ_2}{Q_1^2}\,,\qquad\qquad (\mu,\nu=\pm)\,.
\end{equation}
Thus $K_{\mu\nu}$ is transversal up to the last two terms, but these
correspond to contact terms in coordinate space. Dropping these \lq\lq
seagulls", the coefficient of the transversal part in Fourier space
becomes
\be
\begin{align}
&\tau_l(Q)=-\frac{1}{2}\sum_\omega
\tilde X_\omega^{(l)}(Q)A^{(l)}_\omega\label{taul}\\
&-\frac{1}{2}\sum_{\stackrel{J\equiv l}{\omega}}
\left(\frac{-iM}{2}\right)^J B^{(J,l)}_\omega
\left\{\tilde X^{(J,l)}_\omega(Q)+ 
\tilde{\overline{X}}^{(J,l)}_\omega(Q) 
-4i\pi c_\omega\delta_{J1}\frac{iQ_2}{Q_1^2Q^2}\right\}\,.
\nonumber
\end{align}
\nonumber
\end{equation}
This can alternatively be written as
\be
\begin{align}
\tau_l(Q)&=-\frac{1}{2}\sum_\omega
\tilde X_\omega^{(l)}(Q)A^{(l)}_\omega\label{taul2}
-\frac{1}{2}\sum_{\stackrel{J\equiv l}{\omega}}
\Big\{\left[\left(2MQ_-\right)^J+\left(2MQ_+\right)^J\right]\\
&\left(\frac{{\rm d}}{{\rm d}Q^2}\right)^J
\tilde W^{(J,l)}_\omega(Q^2) B^{(J,l)}_\omega\Big\}+
M\pi\frac{iQ_2}{Q^2(Q^2-Q_2^2)}\delta_{l1}\sum_\omega c_\omega
B^{(1,1)}_\omega, 
\nonumber
\end{align}
\nonumber
\end{equation}
where
\be
W^{(J,l)}_\omega(y^2)=y^2 V^{(J,l)}_\omega(y^2)\,.
\end{equation}
Note that the anomalous contribution to
(\ref{taul2}) is regular on the imaginary $Q_2$ axis 
hence does not contribute to the structure functions. 
Let us also define
\be
\xi^{(J,l)}(y^2)=\sum_\omega W^{(J,l)}_\omega(y^2) B^{(J,l)}_\omega.
\end{equation}
If we now compute
\be
\hat\xi^{(J,l)}(Q^2)=\sum_\omega 
\left(Q^2\right)^{J+1}\left(\frac{{\rm d}}{{\rm d}Q^2}\right)^J
\tilde W^{(J,l)}_\omega(Q^2) B^{(J,l)}_\omega
\end{equation}
we see (using asymptotic freedom) that the coefficient functions
behave as $\left(Q^2\right)^{-t^{(J,l)}_\omega}$, up to logarithmic
corrections. We will keep the contributions of the leading (twist 0)
operators only.
Note that for $J=l=1$ the only twist 0 operator is 
${\cal B}^{(1,1)ab}_1=J^{ab}_+$ with 
\be
B^{(1,1)}_1=4\,.
\end{equation}


From (\ref{taul2}) we obtain for the Taylor coefficients 
\be
\tau_{l;N}(Q^2)=\hat{\tau}_{l;N}(Q^2)-\frac{4\pi 
Mc_1}{Q^2}\,\frac{\delta_{l1}}
{(i\sqrt{Q^2})^{N+1}}\,,
\label{tau1N}
\end{equation}
where (up to higher twist contributions)
\footnote{Note for large $Q^2$ the anomalous term dominates
over $\hat\tau_{l;N}$.}
\be
\hat\tau_{l;N}(Q^2)\cong-\frac{1}{2}\,
\frac{(2M)^N}{\left(Q^2\right)^{N+1}}\,
\hat\xi^{(N,l)}\left(Q^2\right)\,.
\end{equation}
Inserting this in (\ref{w1Ncauchy}) we obtain:
\be
M_{l;N}\left(-Q^2\right)=M\left(\frac{Q^2}{2M}\right)^{N+1}\left\{
\hat 
\tau_{l;N}\left(Q^2\right)+\frac{2M}{Q^2}\delta_{l1}\frac{1}{
(i\sqrt{Q^2})^{N+1}}
\left(1-2\pi c_1\right)\right\}.
\end{equation}
Later we will see that $c_1=1/2\pi$. Thus these two subtle effects cancel
each other, and so the final formula coincides with the naive one:
\be
M_{l;N}\left(-Q^2\right)\cong-\frac{1}{4}\hat\xi^{(N,l)}\left(Q^2\right)\,.
\end{equation}

Up to now we have related the moments to the Taylor coefficients
which we see are determined by the structure of the OPE.
But to get quantitative results at this stage we need more dynamical input. 
This can be supplied by analyzing the OPE in the framework of 
renormalized PT, which is the topic of the next section.

\section{Perturbation theory and operator product expansion}

We consider the $\On$ sigma model Lagrangian
\footnote{In practical computations in infinite volume one usually adds 
a coupling to an external field $-\frac{h_0}{g_0^2}\left(S^n-1\right)$
to serve as an intermediate IR regulator. For IR finite quantities the 
renormalized external field  
$h_{\rm R}=h_0\sqrt{Z}/Z_1$ is set to zero at the end of the computation}
\be
{\cal L}_{\rm E}=\frac{1}{2g_0^2}
\sum_{a=1}^n\,\partial_\mu S^a \,\partial_\mu S^a\,,
\,\,\,\,\,S^2=1\,,
\label{onlag}
\end{equation}
and work in $D=2-\epsilon$ dimensions using dimensional regularization. 
Renormalized fields $S^a_{\rm R}$ and coupling $g$ are given by
\be
\begin{split}
S^a=Z^{1/2}\,S^a_{\rm R}\,,\qquad\ \ 
Z&=1-\frac{\gamma_0g^2}{\epsilon}+\cdots\,,\\
g_0^2=\mu^\epsilon\,g^2\,Z_1\,,\qquad
Z_1&=1-\frac{2\beta_0g^2}{\epsilon}+\cdots\,.
\end{split}
\end{equation}
We denote the usual renormalization group (RG) derivative by
\be
{\cal D}=\mu\,\frac{{\rm d}}{{\rm d}\mu}=\mu\,\frac{\partial}{\partial\mu}
+\beta(g)\,\frac{\partial}{\partial g}\,,
\end{equation}
where the dimensional regularization beta function is
\be
\beta(g)=-\frac{\epsilon}{2}\,g+\bar\beta(g)\,,\qquad
\bar\beta(g)=-\beta_0g^3-\beta_1g^5-\dots
\end{equation}
and
\be
\beta_0=\frac{n-2}{4\pi},\qquad\qquad\qquad
\beta_1=\frac{n-2}{8\pi^2}.
\end{equation}
The RG $\Lambda$--parameter in the $\msbar$ scheme 
satisfies ${\cal D}\Lambda_\msbar=0$ and is written
\be
\Lambda_{\msbar}=\mu\,\rme^{f(g)},
\end{equation}
where
\be
f(g)=-\frac{1}{2\beta_0g^2}-\frac{\beta_1}{2\beta_0^2}
\ln(2\beta_0g^2)+\frac{\gamma}{2}+\rmO(g^2)
\end{equation}
with $\gamma=\ln4\pi+\Gamma^\prime(1)$; (note  
$\frac{\beta_1}{2\beta_0^2}=\frac{1}{n-2}$).

The spin field $\Phi^a$ differs from the renormalized $\On$ field
$S^a_R$ only by a finite renormalization:
\be
\Phi^{a\rm E}=\Omega_n\,S^a_R
\end{equation}
and solving the RG equations for the vacuum two--point function the
standard way we find
\be
\Omega_n^2\,\rme^{-p(g)}=\left(\frac{2\pi}{n-2}\right)^{\frac{n-1}{n-2}}\,
nC_n\,,
\end{equation}
where the constant $C_n$ is that appearing in Eq.~(\ref{Cn})
and $p(g)$ is the solution of
\be
p^\prime(g)=\frac{\gamma(g)}{\bar\beta(g)}\,,
\label{pg}
\end{equation}
where $\gamma(g)$ is the anomalous dimension of the spin field:
\be
\gamma(g)={\cal D}\ln Z=\gamma_0g^2+\cdots\,,\qquad\,
\gamma_0=\frac{n-1}{2\pi}\,.
\end{equation}
The integration constant in (\ref{pg}) is fixed by requiring
\be
\rme^{-p(g)}=\left(g^2\right)^{\frac{\gamma_0}{2\beta_0}}\left\{
1+\rmO(g^2)\right\}\,.
\end{equation}

\subsection{Zero twist operators}

We now introduce a basis for zero twist operators composed of an even 
number of spin fields. For isospin $l=0$ we write
\be
K^{(n_1,m_1)\cdots (n_k,m_k)}=\frac{1}{g_0^{2k}}\,
\left(\partial_+^{n_1}S^{a_1}\cdot \partial_+^{m_1}S^{a_1}\right)
\cdots
\left(\partial_+^{n_k}S^{a_k}\cdot \partial_+^{m_k}S^{a_k}\right),
\end{equation}
where we introduced the notation
\be
\partial_\pm=\frac{1}{2}\left(i\partial_2\mp\partial_1\right)=
\frac{1}{2}(\partial_0\mp\partial_1)\,.
\end{equation}
It is very important to notice that a complete basis can be chosen
such that
\be
n_i,m_i\geq1\,,\,\,\,\,\,i=1,\dots,k\,,
\end{equation}
which can be achieved by using the identity
\be
S^a\,\partial^m_+S^a=-\frac{1}{2}\,\sum_{i=1}^{m-1} \,\binom{m}{i}\,
\partial_+^{m-i}S^a\cdot\partial_+^i S^a\,.
\end{equation}
The spin of the above operators is
$\sum_{i=1}^k\,(n_i+m_i)=J\,,$
whereas the mass dimension is $J-k\epsilon$, i.e. the operators
are of zero twist only in exactly two dimensions. 
For $l=1,2$ we can define the operators 
\be
K^{ab(n_0,m_0)(n_1,m_1)\cdots (n_k,m_k)}=
\frac{1}{g_0^2}\,\partial_+^{n_0}S^a\cdot \partial_+^{m_0}S^b\,
K^{(n_1,m_1)\cdots (n_k,m_k)}
\end{equation}
with spin $\sum_{i=0}^k\,(n_i+m_i)=J\,,$
and dimension $J-(k+1)\epsilon$. For $l=1$ we have to antisymmetrize
the $a,b$ indices, whereas for $l=2$ we take the symmetric, traceless
part. Now
\be
0\leq n_0\leq m_0\,,\qquad\qquad
n_i,m_i\geq1\qquad i=1,\dots,k
\end{equation}
and correspondingly there are three types of operators:

\hspace{3cm}
type I:\ \ $\quad l=0\, \qquad{\rm and}\qquad l=1,2,\quad n_0>0$\,.

\hspace{3cm}
type II:\ $\quad l=1,2,\quad n_0=0,\quad m_0\geq1$\,.

\hspace{3cm}
type III:$\quad l=2, \quad n_0=m_0=0$\,.

It is now straightforward to calculate the potentially divergent
matrix elements of these operators. At one--loop order, after wave
function, charge and mass renormalization we find
\be
\begin{split}
&l=0:\qquad
{\rm all\ matrix\ elements\ finite}\\
&l=1:\qquad
{\rm type\ I:\ \ }\left(1-\frac{g^2}{2\pi\epsilon}\right)
\times {\rm lowest\ order}\\
&\phantom{l=0}\qquad
{\rm type\ II:\ \ only\ type\ I\ operator\ matrix\ elements}\\
&l=2:\qquad
{\rm type\ I:\ \ }\left(1-\frac{g^2}{2\pi\epsilon}\right)
\times {\rm lowest\ order}\\
&\phantom{l=1}\qquad
{\rm type\ II:\ \ }\left(1-\frac{g^2}{\pi\epsilon}\right)
\times {\rm lowest\ order}+{\rm type\ I}\\
&\phantom{l=1}\qquad
{\rm type\ III:\ \ }\left(1-\frac{g^2}{\pi\epsilon}\right)
\times {\rm lowest\ order}+{\rm type\ I,II}
\end{split}\nonumber
\end{equation}

\subsubsection{Operator product expansion at tree level}

The leading terms of the OPE in perturbation theory are simply given by 
Taylor expansion:
\be
\begin{split}
T_{\rm E}&\left(\frac{1}{g_0^2}\,S^a(y)S^b(0)\right)_c=\\
&\qquad
\frac{1}{g_0^2}\,\left(S^aS^b\right)_c+
\frac{1}{g_0^2}\,\sum_{J=1}^\infty\,\frac{1}{J!}\,\left[
\left(\partial_+^JS^a\cdot S^b\right)y_+^J+
\left(\partial_-^JS^a\cdot S^b\right)y_-^J\right]
\end{split}
\end{equation}
up to higher twist operators. The operators appearing in the
sum over $J$ can be written as a sum over operators of definite isospin:
\ba
\frac{1}{g_0^2}\,
\partial_+^JS^a\cdot S^b&=&\sum_{l=0}^2\,{\cal O}^{(J,l)ab}_{(0)}\,,
\\
{\cal O}^{(J,l)ab}_{(0)}&=&\frac{1}{g_0^2}P_l^{ab;cd}
\partial_+^JS^a\cdot S^b\,.
\ea
Important operators with isospin 2 are
\be
\tau^{ab}_{(0)}=\frac{1}{2g_0^2}\left(S^aS^b-\frac{1}{n}\,\delta^{ab}
\right).
\end{equation}
For isospin 1 we have the currents 
\be
J^{ab}_{\mu(0)}=\frac{1}{g_0^2}\left(
S^a\partial_\mu S^b-S^b\partial_\mu S^a\right)\,,
\end{equation}
and for isospin 0 we have the energy--momentum tensor
\be
T_{\mu\nu(0)}=\frac{1}{g_0^2}\left(\partial_\mu
S^a\partial_\nu S^a-\frac{1}{D}\,\delta_{\mu\nu}\,
\partial_\sigma S^a\partial_\sigma S^a\right)\,.
\end{equation}
In terms of these, the leading operators of the OPE can be written as 
\ba
{\cal O}^{(J,2)ab}_{(0)}&=&\partial_+^J\,\tau^{ab}_{(0)}+
{\rm type\ I\ operators}\,,\\
{\cal O}^{(J,1)ab}_{(0)}&=&-\frac{1}{2}
\partial_+^{J-1}\,J^{ab}_{+(0)}+
{\rm type\ I\ operators}\,,\\
{\cal O}^{(2,0)ab}_{(0)}&=&-\frac{\delta^{ab}}{n}\,T_{++(0)}\,.
\ea

\subsubsection{Renormalization of the zero twist operators}

We will now denote by ${\cal B}^{(J,l)ab}_{\alpha(0)}$  the zero twist
operators introduced in the preceding section. Here $\alpha$ is
a multi--index: it includes the operator type I,II,III and possible
further indices. In $D$ space--time dimensions the mass dimension of
${\cal B}^{(J,l)ab}_{\alpha(0)}$ is \break $J-\epsilon\,d^{(J,l)}_\alpha$.
The corresponding renormalized (finite) operators of mass dimension $J$
are:
\be
{\cal B}^{(J,l)ab}_\alpha=\sum_\beta\, Z_{\alpha\beta}^{(J,l)}\,
\mu^{\epsilon d^{(J,l)}_\beta}\,
{\cal B}^{(J,l)ab}_{\beta(0)}\,,
\end{equation}
where the operator renormalization constant matrix is
\be
 Z_{\alpha\beta}^{(J,l)}=\delta_{\alpha\beta}-\frac{g^2}{\epsilon}
w^{(J,l)}_{\alpha\beta}+\cdots.
\end{equation}
We now distinguish the types of operators by writing their multi--indices
\be
\begin{split}
{\cal B}^{(J,l)cd}_{a(0)}&:\ \ {\rm for\ type\ I}\\
{\cal B}^{(J,l)cd}_{\underline{A}(0)}&:\ \ {\rm for\ type\ II}\quad 
(l=1,2)\\
{\cal B}^{(J,2)cd}_{\underline{\underline{A}}(0)}&:\ \ {\rm for\ type\ III}\\
\end{split}
\end{equation}
In this notation the one--loop results of the previous subsection are
\ba
w^{(J,l)}_{ab}&=&\frac{l(l-3)}{4\pi}\,\delta_{ab};\qquad
w^{(J,l)}_{a\underline{A}}=
w^{(J,l)}_{a\underline{\underline{A}}}=0,\\
w^{(J,l)}_{\underline{A}\,\,\underline{B}}&=&\frac{1-l}{\pi}\,
\delta_{AB};\qquad\ \ 
w^{(J,l)}_{\underline{A}\,\,\underline{\underline{B}}}=0,\\
w^{(J,2)}_{\underline{\underline{A}}\,\,\underline{\underline{B}}}&=&
-\frac{1}{\pi}\,\delta_{AB}.
\ea
Using the inverse matrix $W$:
\be
\sum_\beta Z^{(J,l)}_{\alpha\beta}
\, W^{(J,l)}_{\beta\gamma}=\delta_{\alpha\gamma}\,,\qquad
W^{(J,l)}_{\alpha\beta}=\delta_{\alpha\beta}+\frac{g^2}{\epsilon}
w^{(J,l)}_{\alpha\beta}+\cdots
\end{equation}
we define the anomalous dimension matrix
\be
\nu^{(J,l)}_{\rho\sigma}=\sum_\omega\,
Z^{(J,l)}_{\rho\omega}\,\left({\cal D}
W^{(J,l)}_{\omega\sigma}-\epsilon d^{(J,l)}_\omega
W^{(J,l)}_{\omega\sigma}\right),
\end{equation}
which is finite and is given by
\be
\nu^{(J,l)}_{\rho\sigma}=g^2\,w^{(J,l)}_{\rho\sigma}\,(d^{(J,l)}_\sigma-
d^{(J,l)}_\rho-1)+\rmO(g^4)
\end{equation}
to leading order.

We would like to go to a basis where the leading anomalous dimension
matrix is diagonal. This basis is easily found due to the
triangular structure of the leading anomalous dimension matrix.
The renormalized operators in this new basis are denoted 
${\cal B}^{(J,l)cd}_\omega$, where $\omega=a,A$ and ${\cal B}^{(J,l)cd}_a$
are (as before) the renormalized type I operators and 
${\cal B}^{(J,l)cd}_A$ are operators of type II and (for $l=2$) of
type III mixed with lower type operators. In this basis we have
\ba
\nu^{(J,l)}_{ab}&=&\frac{l(3-l)}{4\pi}\,\delta_{ab}\,g^2+
\rmO(g^4)\,,\\
\nu^{(J,l)}_{AB}&=&\frac{l-1}{\pi}\,\delta_{AB}\,g^2+
\rmO(g^4)\,,\qquad(l=1,2)\,,\\
\nu^{(J,l)}_{aA}&=&\nu^{(J,l)}_{Aa}=
\rmO(g^4)\,,\ \quad\qquad\qquad(l=1,2)\,.
\ea

We also note that the canonically dimensionless $l=2$ operator
$\tau^{ab}_{(0)}$ is multiplicatively renormalized: 
\be
\tau^{ab}_{(0)}=Y\,\mu^{-\epsilon}\,\tau^{ab}\,,
\end{equation}
since there is no other operator with the same quantum numbers to mix with.
Here
\be
Y=1-\frac{g^2}{\pi\epsilon}+\cdots
\end{equation}
leading to
\be
{\cal D}\,\ln Y=\frac{g^2}{\pi}+\rmO(g^4)\,.
\end{equation}

It is clear that for the $l=0$ operators we can chose the diagonal
basis so that
\be
{\cal O}^{(J,0)cd}_{(0)}={\cal B}^{(J,0)cd}_{1(0)}
\end{equation}
and for $l=1,2$ the $A=1$ operators so that
\ba
{\cal B}^{(J,2)cd}_{1(0)}=\partial_+^J\,\tau^{cd}_{(0)}=\mu^{-\epsilon}\,
Y\,{\cal B}^{(J,2)cd}_1,\,\,\, &{\rm where}&\,\,\,
{\cal B}^{(J,2)cd}_1=\partial_+^J\,\tau^{cd}\,,
\\
{\cal 
B}^{(J,1)cd}_{1(0)}=\partial_+^{J-1}\,J^{cd}_{+(0)}=\mu^{-\epsilon}\,
{\cal B}^{(J,1)cd}_1,\,\,\, &{\rm where}&\,\,\,
{\cal B}^{(J,1)cd}_1=\partial_+^{J-1}\,J^{cd}_+\,.
\ea
We then have
\ba
{\cal O}^{(J,2)cd}_{(0)}&=&{\cal B}^{(J,2)cd}_{1(0)}+
\sum_{a}\lambda_a^{(J,2)}\,{\cal B}^{(J,2)cd}_{a(0)}\,, 
\\
{\cal O}^{(J,1)cd}_{(0)}&=&-\frac{1}{2}\,{\cal B}^{(J,1)cd}_{1(0)}+
\sum_{a}\lambda_a^{(J,1)}\,{\cal B}^{(J,1)cd}_{a(0)}\,. 
\ea
Finally for $(J,0)=(2,0)$ there is just one operator and we have
\be
{\cal B}^{(2,0)cd}_{1(0)}=-\frac{\delta^{cd}}{n}\,
T_{++(0)}=\mu^{-\epsilon}\,
{\cal B}^{(2,0)cd}_1,\,\,\, {\rm where}\,\,\,
{\cal B}^{(2,0)cd}_1=-\frac{\delta^{cd}}{n}\,T_{++}\,,
\end{equation}
and therefore, the 1--particle matrix element is known exactly:
\be
B^{(2,0)}_1=2\,.
\end{equation}

\subsubsection{The operator product expansion in perturbation theory}

In bare perturbation theory we have
\be
\begin{split}
T_{\rm E}\frac{1}{g_0^2}
&\left(S^a(y)\,S^b(0)\right)_c=\\
&\sum_{l,\omega}{\cal A}^{(l)ab}_{\omega(0)}\,k^{(l)}_{\omega(0)}(y^2)
+\sum_{J=1}^\infty\,\sum_{l,\omega}
\,k^{(J,l)}_{\omega(0)}(y^2)\left\{ {\cal B}^{(J,l)ab}_{\omega(0)}\,y_+^J
+\overline{{\cal B}}^{(J,l)ab}_{\omega(0)}\,y_-^J\right\}\,,
\end{split}
\end{equation}
which, after renormalization, becomes
\be
\begin{split}
T_{\rm E}&\left(S^a_R(y)\,S^b_R(0)\right)_c=\\
&\sum_{l,\omega}{\cal A}^{(l)ab}_\omega\,k^{(l)}_\omega(y^2)
+\sum_{J=1}^\infty\,\sum_{l,\omega}
\,k^{(J,l)}_\omega(y^2)\left\{ {\cal B}^{(J,l)ab}_\omega\,y_+^J
+\overline{{\cal B}}^{(J,l)ab}_\omega\,y_-^J\right\}\,,
\end{split}
\end{equation}
where
\be
k^{(J,l)}_\omega(y^2)=\mu^\epsilon\,\frac{g^2Z_1}{Z}\,\sum_\rho
k^{(J,l)}_{\rho(0)}(y^2)\,\mu^{-\epsilon\,d^{(J,l)}_\rho}\,W^{(J,l)}_{
\rho\omega}
\end{equation}
which satisfies the renormalization group equation (RGE)
\be
({\cal D}+\gamma(g))\,k^{(J,l)}_\sigma=\sum_\omega k^{(J,l)}_\omega\,
\nu^{(J,l)}_{\omega\sigma}\,.
\end{equation}
Finally the original coefficient functions $\gamma^{(J,l)}_\omega$ of 
(\ref{OPE}) are related to the renormalized coefficients $k^{(J,l)}_\omega$
by
\be
\gamma^{(J,l)}_\omega=\Omega_n^2\,k^{(J,l)}_\omega\,,
\end{equation}
and so (\ref{eta}) can now be written as
\be
\eta^{(J,l)}(y^2)=\Omega_n^2\,\sum_\omega\,k^{(J,l)}_\omega(y^2)\,
B^{(J,l)}_\omega\,.
\end{equation}

The perturbative expansion of the renormalized coefficient functions is
\be
k^{(J,l)}_\omega(\mu|y|,g)=g^2\,K^{(J,l)}_\omega
+g^4\,\tilde q^{(J,l)}_\omega(\mu|y|)+\rmO(g^6)\,.
\end{equation}
We already computed the leading (tree--level) terms:
\ba
K^{(J,0)}_\omega&=&\frac{1}{J!}\,\delta_{\omega1}\,,\\
K^{(J,1)}_A&=&-\frac{1}{2J!}\,\delta_{A1}\,,
\qquad\qquad K^{(J,1)}_a=\frac{1}{J!}\,\lambda^{(J,1)}_a\,,\\
K^{(J,2)}_A&=&\frac{1}{J!}\,\delta_{A1}\,,
\qquad\qquad\quad\  K^{(J,2)}_a=\frac{1}{J!}\,\lambda^{(J,2)}_a\,.
\ea

A building block used in the solution of the RGE is the matrix
$\hat U^{(J,l)}_{\omega\sigma}(g)$, which is a solution of the matrix
differential equation
\be
\bar\beta(g)\,\frac{\partial}{\partial g}\,\hat U^{(J,l)}_{\omega\sigma}(g)=
-\sum_\rho\,\nu^{(J,l)}_{\omega\rho}(g)\,\hat U^{(J,l)}_{\rho\sigma}(g).
\label{hatU}
\end{equation}
If we have such a solution and its matrix inverse $U$ satisfying
\be
\sum_\rho U^{(J,l)}_{\omega\rho}(g) \hat U^{(J,l)}_{\rho\sigma}(g)=
\delta_{\omega\sigma}\,,
\end{equation}
we can build the RG invariant coefficient
\be
G^{(J,l)}_\omega=\rme^{p(\bar g)}\,\sum_\rho\, k^{(J,l)}_\rho(1,\bar g)\,
\hat U^{(J,l)}_{\rho\omega}(\bar g)
\label{Gamma}
\end{equation}
and the RG invariant numbers
\be
V^{(J,l)}_\omega=\sum_\rho\,U^{(J,l)}_{\omega\rho}(g)\,B^{(J,l)}_\rho\,.
\label{numbers}
\end{equation}
In (\ref{Gamma}) the running coupling $\bar g$ is defined as the solution 
of
\be
f(\bar g)=f(g)+\ln(\mu|y|)\,,
\end{equation}
which, for small $|y|$, has the asymptotic expansion
\be
2\beta_0\bar g^2=\lambdat+c\lambdat^2+\rmO(\lambdat^3)
\,,\qquad c=\frac12(\Gamma'(1)-\ln\pi)\,,
\end{equation}
where the effective coupling $\lambdat$ is defined by
\be
\frac{1}{\lambdat}+\frac{1}{n-2}\,\ln\lambdat
=\ln\frac{2\rme^{\Gamma^\prime(1)}}{\Lambda_\msbar\vert y\vert}\,.
\label{lambdat}
\end{equation}

Putting the building blocks together we have
\be
\eta^{(J,l)}(y^2)=\left(\frac{2\pi}{n-2}\right)^{\frac{n-1}{n-2}}
\,n\,C_n\sum_\omega\,G^{(J,l)}_\omega\,V^{(J,l)}_\omega\,.
\end{equation}
Note that if with some $\hat Y$
\be
{\cal B}^{(J,l)cd}_{1(0)}=\mu^{-\epsilon}\,\hat Y\,{\cal B}^{(J,l)cd}_1\,,
\end{equation}
which is the case for $l=1,2$ and also for $(J,l)=(2,0)$, then
\be
\nu^{(J,l)}_{1\omega}={\cal D}\ln \hat Y\,\delta_{1\omega}\,.
\label{kisy}
\end{equation}

\subsubsection{Solution of the matrix problem}

In this subsection we will omit the upper index $^{(J,l)}$ and use
matrix notation. We want to solve
\be
\bar\beta(g)\,\frac{\partial}{\partial g}\,\hat U(g)=
-\nu(g)\,\hat U(g),
\label{RGE}
\end{equation}
which is (\ref{hatU}) in this notation. 
We know that in our basis
\be
\nu(g)=2\beta_0\,\Delta\,g^2+\rmO(g^4),
\end{equation}
where $\Delta$ is a diagonal matrix with diagonal elements
\be
\begin{split}
l&=0:\qquad\qquad \Delta_a=0,\\
l&=1:\qquad\qquad \Delta_A=0,\qquad\qquad \ \ \,\Delta_a=
\frac{1}{n-2}\,,\\
l&=2:\qquad\qquad \Delta_A=\frac{2}{n-2},\qquad\qquad \Delta_a=
\frac{1}{n-2}\,.\\
\end{split}
\end{equation}
Using the expansion
\be
\frac{\nu(g)}{\bar\beta(g)}=-\frac{2\Delta}{g}-2\,\sum_{p=1}^\infty\,
g^{2p-1}\,A^{(p)}\,,
\end{equation}
we can take the Ansatz
\be
\hat U(g)=[1+R(g)]\,g^{2\Delta}
\end{equation}
with
\be
R(g)=\sum_{s=1}^\infty\,g^{2s}\,R^{(s)}\,,
\end{equation}
and put it into (\ref{RGE}). We get
\be
s\,R^{(s)}+[R^{(s)},\Delta]=A^{(s)}+\sum_{p=1}^{s-1}\,A^{(s-p)}\,R^{(p)},
\qquad\qquad s=1,2,\dots,
\end{equation}
which has a unique recursive solution unless $\Delta_\omega-\Delta_\sigma=s$
occurs for some $\omega,\sigma$ and $s$. In our case this is possible only
for $s=1$ and only if $n=3$. For $n=3$ we thus take the modified Ansatz
\be
\hat U(g)=[1+R(g)+\ln g^2\,\tilde R(g)]\,g^{2\Delta}
\end{equation}
with
\be
R(g)=\sum_{s=1}^\infty\,g^{2s}\,R^{(s)},\qquad\qquad
\tilde R(g)=\sum_{s=1}^\infty\,g^{2s}\,\tilde R^{(s)}.
\end{equation}
In this case we start with
\begin{align}
l&=1: & \tilde R^{(1)}_{aA}&=A^{(1)}_{aA}\,, &\tilde R^{(1)}_{ab}&=0\,,
& \tilde R^{(1)}_{Aa}&=0\,,& \tilde R^{(1)}_{AB}&=0\,,\\
& & R^{(1)}_{aA}&=0\,,& R^{(1)}_{ab}&=A^{(1)}_{ab}\,,&
R^{(1)}_{Aa}&=\frac{1}{2}\,A^{(1)}_{Aa}\,,&
R^{(1)}_{AB}&=A^{(1)}_{AB}\,,
\end{align}
and
\begin{align}
l&=2: & \tilde R^{(1)}_{Aa}&=A^{(1)}_{Aa}\,,&\tilde R^{(1)}_{AB}&=0\,,
& \tilde R^{(1)}_{ab}&=0\,,& \tilde R^{(1)}_{aA}&=0\,,\\
& & R^{(1)}_{Aa}&=0\,,& R^{(1)}_{AB}&=A^{(1)}_{AB}\,,&
R^{(1)}_{ab}&=A^{(1)}_{ab}\,,&
R^{(1)}_{aA}&=\frac{1}{2}\,A^{(1)}_{aA}\,,
\end{align}
and after that there is a unique, recursive solution of the system
\ba
s\,\tilde R^{(s)}+[\tilde R^{(s)},\Delta]&=&
\sum_{p=1}^{s-1}\,A^{(s-p)}\,\tilde R^{(p)}\,,\\
s\, R^{(s)}+[ R^{(s)},\Delta]&=&A^{(s)}-\tilde R^{(s)}+
\sum_{p=1}^{s-1}\,A^{(s-p)}\, R^{(p)}
\ea
for $s=2,3,\dots$

Note that from the recursion relations it follows that
\ba
\tilde R^{(s)}_{\omega a}&=&0\qquad {\rm for\ \ }l=1\,,
\\
\tilde R^{(s)}_{\omega A}&=&0\qquad {\rm for\ \ }l=2\,.
\ea
We also note that because $\nu_{1\omega}$ is proportional to $\delta_{1\omega}$
\be
\tilde R^{(s)}_{1\omega}=0 \qquad\qquad {\rm and} \qquad\qquad
R^{(s)}_{1\omega}\sim\delta_{1\omega}
\end{equation}
and therefore
\be
\hat U_{1\omega}=\frac{1}{w}\,\delta_{1\omega}\qquad\qquad
U_{1\omega}=w\delta_{1\omega}\,,
\end{equation}
where $w$ is the solution of
\be
\bar\beta(g)w^\prime(g)=y(g)w(g)\,,
\end{equation}
where $y(g)={\cal D}\ln\hat Y(g)$, the coefficient occuring in Eq. 
(\ref{kisy}).
This has the following important consequences.
\begin{align}
V_1&=0 &{\rm for}\,\,l=2\,,\,\,\,\,\,{\rm and\,\,for}\,\,l=1,\ J>1\,,\\
V_1&=B_1 &{\rm for}\,\,l=1,\ J=1\,,\,\,\,\,\,{\rm and\,\,for}\,\,l=0,\ J=2\,.
\end{align}

\subsubsection{Leading terms in coordinate space}

Using the results of the preceding subsections we can calculate the
leading terms in the short distance expansion of the functions
(\ref{eta}). We find
\ba
\eta^{(J,0)}&=&\frac{1}{J!}\,f^{(J)}_0\lambdat^{-\frac{1}{n-2}}\left\{
1+\rmO(\lambdat)\right\}\,,
\\
\eta^{(1,1)}&=&f^{(1)}_1\lambdat^{-\frac{1}{n-2}}\left\{
1+\rmO(\tilde \lambda)\right\}\,,
\ea
where
\ba
f^{(J)}_0&=&\frac{2\pi n\,C_n}{n-2}\,V_1^{(J,0)}\,,
\\
f^{(1)}_1&=&-\frac{\pi n\,C_n}{n-2}\,B_1^{(1,1)}\,.
\ea

For the case $l=1$, $J>1$ we have to distinguish between the cases
$n>3$ and $n=3$. In the former case
\be
\eta^{(J,1)}={\rm const.}+\frac{1}{J!}\,f^{(J)}_1\,
\lambdat^{\frac{n-3}{n-2}}\,
\left\{1+\rmO(\lambda^{\frac{1}{n-2}})\right\},
\end{equation}
where
\be
f^{(J)}_1=
\frac{4\pi^2 n\,C_n}{(n-2)^2}\,\sum_A\,{\cal L}^{(J)}_A\,V^{(J,1)}_A,
\label{tilW}
\end{equation}
and
\be
{\cal L}^{(J)}_A=-\frac{1}{2}\,R^{(1)(J,1)}_{1A}+
\sum_{a}\lambda^{(J,1)}_a\,R^{(1)(J,1)}_{aA}+J!\,\tilde q^{(J,1)}_A.
\end{equation}
In the $n=3$ case we have
\be
\eta^{(J,1)}={\rm const.}+\frac{1}{J!}\,f^{(J)}_1\,\ln\tilde \lambda\,
\left\{1+\rmO(\tilde \lambda)\right\}\,,
\end{equation}
where
\be
f^{(J)}_1=\frac{4}{\pi}\,\sum_A\,\hat{\cal L}^{(J)}_A\,V^{(J,1)}_A\,,
\qquad\qquad
\hat{\cal L}^{(J)}_A=
\sum_{a}\lambda^{(J,1)}_a\,\tilde R^{(1)(J,1)}_{aA}\,.
\end{equation}

Finally for $l=2$ we find
\be
\eta^{(J,2)}={\rm const.}+\frac{1}{J!}\,f^{(J)}_2\,\tilde \lambda\,
\left\{1+\rmO(\lambdat^{\frac{1}{n-2}})\right\}\,,
\end{equation}
where
\be
f^{(J)}_2=\left(\frac{2\pi}{n-2}\right)^{\frac{2n-3}{n-2}}n\,C_n
\,\sum_a\,{\cal K}^{(J)}_a\,V^{(J,2)}_a\,,
\end{equation}
and
\be
{\cal K}^{(J)}_a=R^{(1)(J,2)}_{1a}+
\sum_b\lambda^{(J,2)}_b\,R^{(1)(J,2)}_{ba}+J!\,\tilde q^{(J,2)}_a\,.
\end{equation}

Using the coordinate space results above
and the asymptotic formulae of Appendix~\ref{Asy}, we are now
in a position to derive the results on leading large momentum behavior of
the spin structure function moments given in subsection~4.1.

\subsection{OPE for the currents}

It is straightforward to calculate the leading operator product coefficients
in perturbation theory:
\be
\begin{align}
W^{(J,0)}_{\omega (0)}(y^2)&=-\frac{2}{J!}\,\frac{1}{(n-1)g_0^2}
\delta_{\omega1}+\rmO(1)\,,\qquad\qquad J\geq2\label{pt1}\\
W^{(J,1)}_{\omega (0)}(y^2)&=\frac{1}{2J!}\,\frac{1}{(n-2)g_0^2}
\delta_{\omega k}+\rmO(1)\,,\qquad\qquad \,\,J\geq3\label{pt2}\\
{\cal Y}W^{(1,1)}_{1 (0)}(y^2)
&=-\frac{1}{8\pi}\left(1-\frac{g_0^2}{2\pi}\right)+
\rmO\left(g_0^4\right)\,.\label{pt3}
\end{align}
\nonumber
\end{equation}
where the operator associated with (\ref{pt2}) is
\be
{\cal B}^{(J,1)ab}_{k(0)}=\frac{1}{g_0^2}\left(
\partial_+^{J-1}S^a\cdot\partial_+S^b-
\partial_+^{J-1}S^b\cdot\partial_+S^a\right)\,.
\end{equation}
Eqns.~(\ref{pt1}) and (\ref{pt2}) can be obtained by tree--level 
perturbation theory, 
while the results necessary to write down the one--loop formula 
(\ref{pt3}) can be found in \cite{CMP}. Also the results of 
\cite{Luscher,CMP} show that 
the coefficient in (\ref{tau1N}) is given by
\be
c_1=\frac{1}{2\pi}\,.
\end{equation}

Using renormalization group improved perturbation theory we can write
\be
\xi^{(J,l)}(y^2)=\sum_\omega\Gamma^{(J,l)}_\omega(y^2)V^{(J,l)}_\omega,
\label{xiRG}
\end{equation}
where, as in (\ref{numbers}),
\be
V^{(J,l)}_\omega=\sum_\rho U^{(J,l)}_{\omega\rho}(g)\, B^{(J,l)}_\rho
\label{Vsl}
\end{equation}
are renormalization group invariant constants and
\be
\Gamma^{(J,l)}_\omega(y^2)=\sum_\rho W^{(J,l)}_\rho(y^2)\,
\hat U^{(J,l)}_{\rho\omega}(\bar g)
\label{Gammasl}
\end{equation}
are renormalization group invariant coefficient functions.
Putting everything together we arrive at the results already given 
in subsect.~4.1.

\section{Structure functions for $n=3$}

In this section we consider the case $n=3$,
where it is possible to compute the structure functions
accurately in the whole range of $x$ for a given $q^2$. 
The case $n=3$ is rather special for various reasons 
e.g. the spin and
current 2--point functions exhibit in this case very similar properties
and there are miraculous scaling relations \cite{JanosMax} which relate
them \footnote{see also the OPE in sect.~4}.
In the S--matrix bootstrap approach its distinguishing feature is 
that it is the model for which the $r$--particle form factors
can most easily be obtained explicitly. They take the simple form
\be
f^a_{b_1\dots b_r}(\theta_1,\dots,\theta_r)=\Psi_r(\theta_1,\dots,\theta_r)
g^a_{b_1\dots b_r}(\theta_1,\dots,\theta_r)\,,
\label{reducedff}
\end{equation}
where
\ba
\Psi_r(\theta_1,\dots,\theta_r)&=&\frac12\pi^{3r/2-1}
\prod_{1\le i<j\le r}\psi(\theta_i-\theta_j)\,,
\\
\psi(\theta)&=&\frac{\theta-\pi i}{\theta(2\pi i-\theta)}
{\tanh^2\frac{\theta}{2}}\,,
\ea
and the reduced form factors $g^a_{b_1\dots b_r}$ are polynomials in the 
rapidities. There are well defined recursive
procedures for computing the form factors, the only
practical limitation being that they become very involved.
So far the record we have achieved is the 7--particle form factor
of the spin field \cite{JanosPeter1}; already its algebraic expression in
MAPLE involves many megabytes of storage. Fortunately for the structure
functions we only require sums over bilinear factors of the form factors
which are computationally more manageable. In correspondence to 
(\ref{reducedff}) we define reduced form factor squares 
$j_l^{(r)}$ through
\ba
J_l^{(r)}(\bar{\beta}_1,\dots,\bar{\beta}_r)
&=&|\Psi_{r+1}(i\pi,\bar{\beta}_1,\dots,\bar{\beta}_r)|^2
j_l^{(r)}(\bar{\beta}_1,\dots,\bar{\beta}_r)
\\
&=&\frac14\pi^{3r+1}
\left[\prod_{i=1}^rA(\bar{\beta}_i)\right]
\left[\prod_{1\le j<k\le r}B(u_{jk})\right]
j_l^{(r)}(\bar{\beta}_1,\dots,\bar{\beta}_r)\,,
\ea
where
we have introduced two new functions
\ba
A(\theta)&\equiv&\psi(i\pi -\theta)^2=
\frac{\theta^2}{(\theta^2+\pi^2)^2}\frac{1}{\tanh^{4}\frac{\theta}{2}}\,,
\\
B(\theta)&\equiv&|\psi(\theta)|^2=
\frac{\theta^2+\pi^2}{\theta^2(\theta^2+4\pi^2)}\tanh^{4}\frac{\theta}{2}\,.
\ea
The reduced form factor squares for $r=2,3,4$ are given in Appendix 
\ref{apph}.
For $r>4$ the expressions are too lengthy to exhibit in print; the
results for $r=5,6$ can be obtained in the form of files from the authors.
 
For $r=1$  we then have (noting that $\sinh \frac12 b=\kappa$ for $x=1$),
\be 
w_l^{(1)}=\frac{m_l\pi^4}{4}A(b)\delta(1-x)\,,
\end{equation}
where the factors $m_l$ are given in (\ref{ml}).

\subsection{Case $r=2$}

For the case $r=2$ the delta--function constraint in the integral
representation is simply solved
and we obtain the analytic expression 
\be
w_l^{(2)}(q^2,x)=\theta(\omega-2)
\frac{\pi^6\kappa^2}{8\omega\sqrt{\omega^2-4}}
A(\Lambda+\phi/2)A(\Lambda-\phi/2)B(\phi){\cal C}_l(\Lambda,\phi)\,,
\end{equation}
where the kinematic variables $\omega,\Lambda,\phi$ are given by
\ba
2\cosh \frac{\phi}{2}&=&\omega\,,\,\,\,
\label{kinphi}
\\
\omega&=&W/M\,,\,\,\,\,\,W^2=(p+q)^2\,,
\\
\cosh\Lambda&=&(M^2+pq)/(MW)\,,
\ea
and
\be
{\cal C}_l(\Lambda,\phi)=j_l^{(2)}(\Lambda+\phi/2,\Lambda-\phi/2)\,.
\end{equation}
Using the expression for $j_l^{(2)}$ in (\ref{j0})-(\ref{j2}) we have,
\ba
{\cal C}_0&=&\phantom{-}8\pi^2+4\Lambda^2+3\phi^2\,,
\\
{\cal C}_1&=&-2\pi^2+2\Lambda^2+\frac12\phi^2\,,
\\
{\cal C}_2&=&\phantom{-}2\pi^2-2\Lambda^2+\frac32\phi^2\,.
\ea

Despite its relative simplicity, this case exhibits many features
in common with higher $r$.
The structure function approaches its asymptotic values 
very slowly e.g. for $q^2$ fixed 
\be
\omega_l^{(2)}(q^2,x)\sim e_l
\frac{2\pi^2}{\left(1+\frac{1}{4\kappa^2}\right)^2}
\frac{1}{x\ln^2\left(\frac{4\kappa^2}{x}\right)}\,\,\,\,
{\rm for}\,\,\,x\to0\,,
\end{equation}
with $e_0=1,\,e_1=e_2=1/4$ (consistent with the small $x$ behavior
derived in sect.~3), while for $-q^2\to\infty\,,x$ fixed we have
\be
\omega^{(2)}_l\sim e_l\frac{\pi^6 xA(-\ln(1-x))}{8(1-x)
(\ln(-q^2/M^2))^2}\,.
\end{equation}

\subsection{Results for the entire $x$ range}

\begin{figure}[t]
\begin{center}
\hspace{1.0cm}
\epsfig{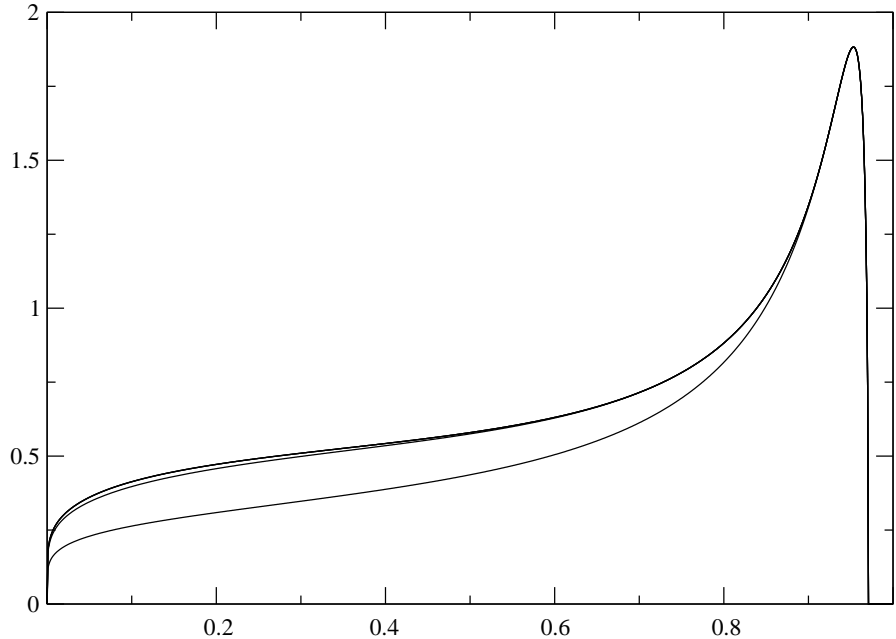}
\caption{\footnotesize Approximations to
$x(w_0+\wt_0)$ as functions of $0<x<1$ for $-q^2/M^2=100$.
Curves correspond to sums up to and including $2,3,4,5,6$--particle
intermediate states. The last 3 curves are indistinguishable on this
scale.
}
\label{wwtilde100}
\end{center}
\end{figure}

\begin{figure}[t]
\begin{center}
\hspace{1.0cm}
\epsfig{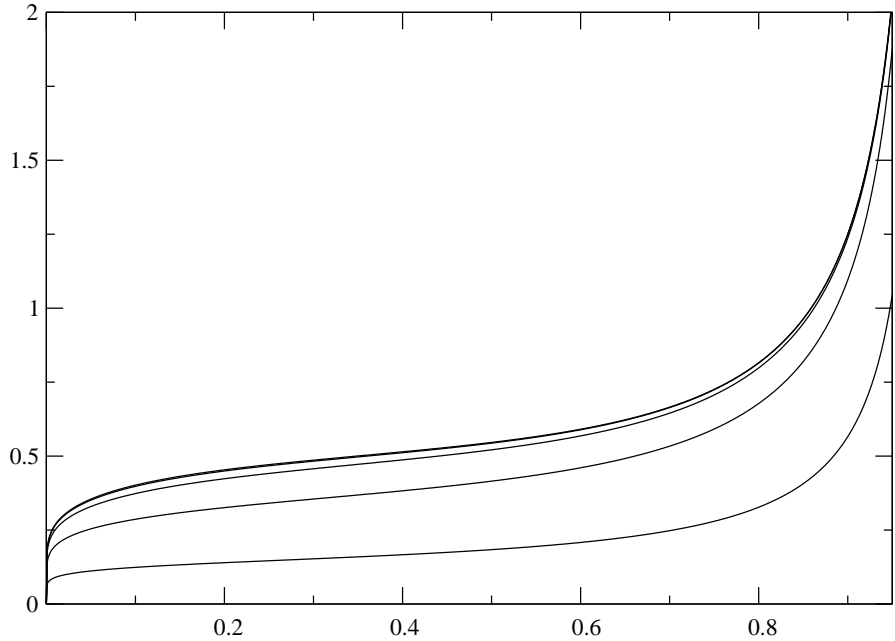}
\label{wwtilde10000}
\caption{\footnotesize Approximations to
$x(w_0+\wt_0)$ as functions of $0<x\le0.95$ for $-q^2/M^2=10^4$.
Curves correspond to sums up to and including 2,3,4,5,6--particle states.
}
\end{center}
\end{figure}

Just as for the 2--point functions \cite{JanosMax}
we find that for a fixed $-q^2$ only
states with a limited number of particles contribute significantly.
To appreciate this better we consider the sum of the field and current
structure functions, which is a rather peculiar thing to do
in general, but which is in fact rather natural in the special case $n=3$.
Figs.~1 and 2 illustrate how the structure
function $x(w_0+\wt_0)$ is built up from
states with increasing particle number for the cases $-q^2/M^2=10^2$ and
$-q^2/M^2=10^4$ respectively. We see that the higher states
contribute very little and
that we obtain nearly exact values for the structure functions
for all values of $-q^2/M^2<10^5$ by including only
intermediate states with $\le 5$ particles for the current
and $\le 6$ particles for the spin field.

\begin{figure}[t]
\begin{center}
hspace{1.0cm}
\epsfig{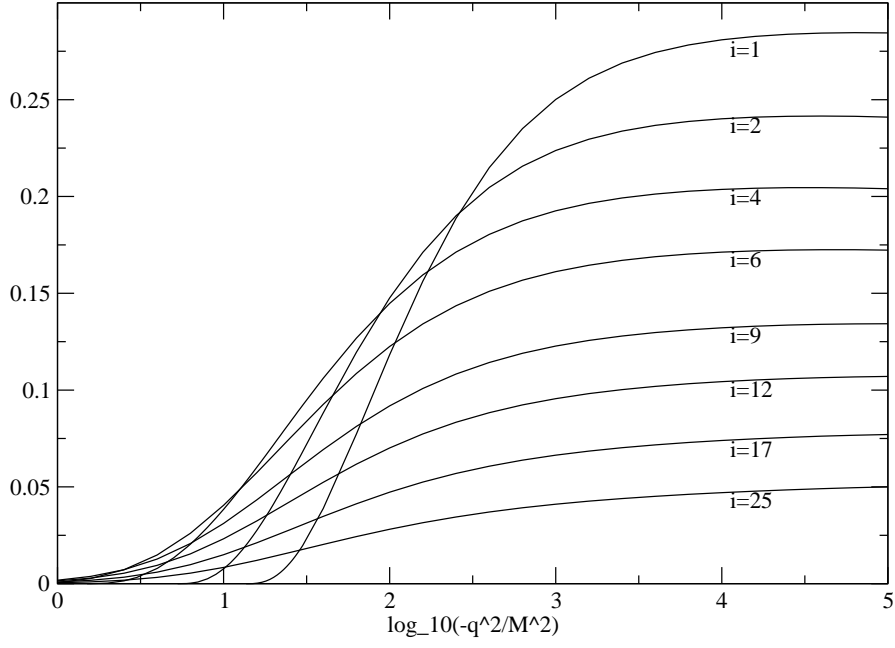}
\label{xw0HERA}
\caption{\footnotesize $xw_0(q^2,x)$ for various values of $x=10^{-i/5}$.
}
\end{center}
\end{figure}
In Fig.~3 we plot $xw_0(q^2,x)$ as a function
of ${\rm log}_{10}(-q^2/M^2)$, for a selection of $x$--values
\footnote{For this model we prefer to show this rather
than the typical HERA plot
where one adds $-\log_{10}(x)$ to separate the $x$--values,
because the latter would obscure the $-q^2$ variation which is rather
small compared to the variation of $-\log_{10}(x)$.}.
The function increases as $-q^2$ increases for all 
values of $x$ in this range and seems consistent with Bjorken scaling
as mentioned in sect.~4.

\subsection{Threshold behavior}  

Note that in Fig.~2 we have cut off the plot at
$x=0.95$. This is because near $x=1$ the function develops a big bump
with a peak $\sim 70$ which, if included in the same plot, would
obscure the features we wanted to show there.
The behavior of the $\sigma$--model structure functions
near $x=1$ is indeed rather involved. For a fixed $-q^2$
the contribution to the structure function
from the $r$--particle state $w^{(r)}$ vanishes for $x$ greater
than some threshold value 
\be
x_r(-q^2)=\left[1-(r^2-1)M^2/q^2\right]^{-1}\,.
\end{equation}
The big bump in $x(w_0+\wt_0)$ referred to above is at this value of
$-q^2/M^2=10^4$ practically entirely due to the $2$--particle
contribution. For this contribution:
\ba
w^{(2)}_l&\sim& E_l(q^2)\sqrt{x_2(-q^2)-x}\,,\,\,\,x\to x_2\,,\,\,\,-q^2
\,\,\,\,{\rm fixed}\,,
\\
w^{(2)}_l&\sim& F_l(x)/\ln^2(-q^2/M^2)\,,\,\,\,-q^2\to\infty\,,\,\,\,x
\,\,\,\,{\rm fixed}\,,
\ea
where $E_l,F_l$ are some (known) functions. The bump arises because
$F_l$ is quite singular near threshold, $F_l\sim [(1-x)\ln^2(1-x)]^{-1}$.
The analytic behavior as $x\to x_2$ sets in only extremely close
to threshold e.g for $-q^2/M^2=10^4$
the position of the peak of the bump is
at $x=0.99954$ whereas the function vanishes at $x_2=0.99970$.
At $-q^2/M^2=10^4$ the 3--particle contribution also has a bump
but it is less pronounced; (peak value $\sim 2.5$ at $x\sim0.9953$).
We conjecture that the threshold behavior
of $w_0^{(r)}$ in the O(3) model is $(x_r-x)^{(r^2-3)/2}$.

\subsection{Moments}

For $r=1$ the moments (\ref{fixmoments}) are simply given by
\be
M_{l;N}^{(1)}(q^2)=\frac{m_l\pi^4}{4}A(2\,{\rm asinh}\kappa)\,.
\end{equation}
The moments with $N>1$ are quite simple to evaluate numerically, 
but for $N=1$ some care must be taken to obtain accurate results.

The problem arises already for $r=2$ where we have integration 
just over $u_1$. 
The $\psi$-factor together with the $x^2$ factor in the integrand is 
\be
F_\psi^{(2)}=A(\bar{\beta}_1)A(\bar{\beta}_2)B(u_1)\bar{x}^2\,.
\end{equation} 
Now for $u_1$ very large  
$\bar{\beta}_2\sim(-q^2+M^2)\rme^{-u_1}$ is exponentially small
and so also $\bar{x}\sim -q^2\rme^{-u_1}$.
Noting i) for $u_1$ large $\bar{\beta}_1\sim u_1$ and ii)
$A(\theta)\sim 16/(\pi^4\theta^2)$ for $\theta\to0$ we have
for large $u_1$ 
\be
F_\psi^{(2)}\sim \left(\frac{-4q^2}{\pi^2(-q^2+M^2)}\right)^2 A_B(u_1)
\label{Fpsi2}
\end{equation}
where 
\be
A_B(u)=A(u)B(u)=\frac{1}{(u^2+\pi^2)(u^2+4\pi^2)}\,, 
\end{equation}
The integral over large $u_1$ gives a sizeable contribution because the 
integrand decays only as $u_1^{-2}$. The integral is thus broken up
into two parts where for the large $u_1$ region the substitution
(\ref{Fpsi2}) is made and there computation of exponential
functions of large argument are not necessary. 

For the case of higher $r$ the procedure is similar. Here the
integrations over $u_1,\dots,u_{r-2}$ can be done safely by introducing for 
them (large) cutoffs (and monitoring the dependence on them), since the
integrands are exponentially suppressed. But for the integration over 
large $u_{r-1}$  the integrand is not exponentially suppressed and in 
this region one replaces the corresponding $\psi$--factor by
\be
F_\psi^{(r)}\sim
\left(\frac{-4q^2}{\pi^2\left[-q^2+M^{(r-1)}({\bf u})^2\right]}\right)^2
\cdot\prod_{j=1}^{r-1}A_B(u_{jr})
\prod_{1\le k<l\le r-1}B(u_{kl})\,.
\end{equation}

\begin{figure}[t]
\begin{center}
\hspace{1.0cm}
\epsfig{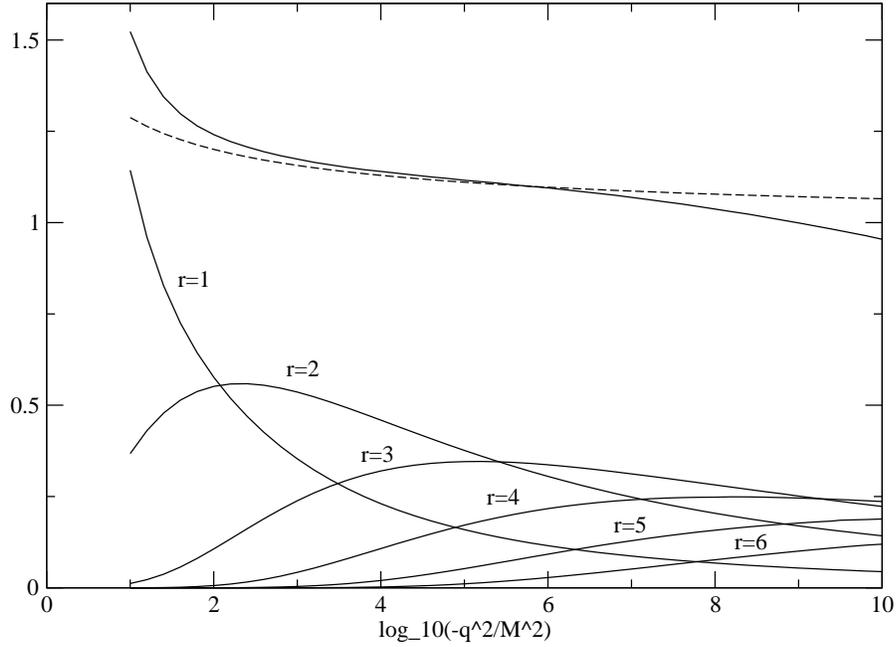}
\caption{\footnotesize
Contributions $M_{0;2}^{(r)}$ for $n=3$ from $r=1,\dots,6$--particle
states. The upper full line is their sum.
The dashed line is the perturbative expansion
of $M_{0;2}+\Mt_{0;2}=1+\lambda$
up to and including terms of order $\lambda(q^2)$.
}
\end{center}
\label{FigM02}
\end{figure}

\begin{figure}[t]
\begin{center}
\hspace{1.0cm}
\epsfig{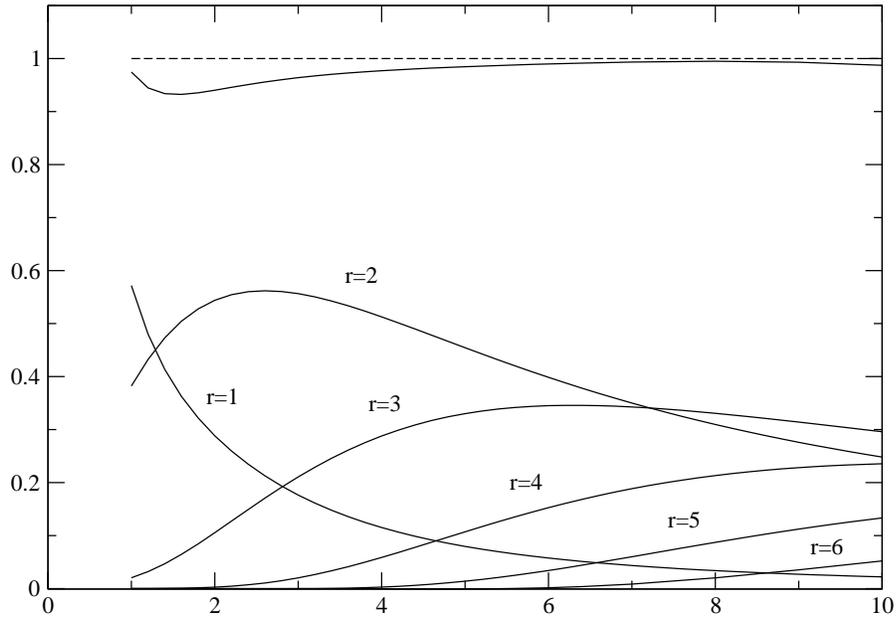}
\caption{\footnotesize
As for Fig.~4 but for the moment $l=1,N=1$; here the
PT result is $1+\rmO(\lambda^2)$.}
\end{center}
\label{FigM11}
\end{figure}

In Figs.~4 and 5 we plot the separate $r$--particle
contributions $M_{0;2}^{(r)}$ and $M_{1;1}^{(r)}$ respectively;
some corresponding numbers are given in 
tables~\ref{mkrmomentsl0}--\ref{mkrmomentstot} in 
Appendix~\ref{momtables}. They are typically bell--shaped
(except for $r=1$) and perhaps obey scaling relations
similar to those of the spectral functions examined in
ref.~\cite{JanosMax}. The figures show how they build up
the sum of moments $M_{0;2}+\Mt_{0;2}$ and $M_{1;1}+\Mt_{1;1}$.
Using the exact ratio of the mass to the $\Lambda$--parameter 
Eq.~(\ref{Lambdaoverm}), we also exhibit the perturbative results up to 
and including terms of order $\lambda(q^2)$. The agreement of the
summed terms and PT is impressive for $-q^2/M^2\sim 10^5$.
For values of $-q^2/M^2>\sim 10^6$ contributions from states with
$\ge 7$ particles must be taken into account. Note we have also
included the contribution of the one particle states in the sums;
these tend to improve the agreement at lower values of $-q^2$
and fall asymptotically as $M^{(1)}_{l;N}\sim 
m_l\pi^4/[4\ln^2(-q^2/M^2)]$.

\section{Sigma model structure functions in the $1/n$ approximation}

\subsection{The spin field structure functions}

In the framework of the $1/n$ approximation the spin amplitude
(\ref{Sigma}) has an expansion of the form 
\be
\Sigma^{ab;cd}(p,q)=\sum_{r=1}^\infty 
\frac{1}{n^r}
\left[S_1^{[r]}\delta^{ac}\delta^{bd}
+S_2^{[r]}\delta^{ad}\delta^{bc}+S_3^{[r]}\delta^{ab}\delta^{cd}\right]\,,
\end{equation}
and so for the various isospin channels
\be
\wt_l(q^2,x)=\sum_{r=1}^\infty\frac{1}{n^r}\wt_l^{[r]}(q^2,x)\,,
\label{wt1on}
\end{equation}
with
\ba
\wt_0^{[r]}&=&S_1^{[r]}+S_2^{[r]}+S_3^{[r+1]}\,,
\label{wt0r}
\\
\wt_1^{[r]}&=&S_1^{[r]}-S_2^{[r]}\,,
\\
\wt_2^{[r]}&=&S_1^{[r]}+S_2^{[r]}\,.
\ea

The Feynman rules for the $1/n$ expansion of the $\sigma$--model
has been described in many places (see e.g. ref.~\cite{largenref}
and references therein) and will not be repeated here. 
We only mention that the diagrams involve 
the bare propagator of the fundamental spin field, 
and the bare propagator $B^{-1}$ of an auxiliary
isospin scalar composite field, which we call $\lambda$, 
which is the inverse
of the scalar 1--loop integral given in App.~\ref{app1on}.

In leading order $1/n$ the only contribution 
to the scalar structure function is the tree diagram involving  
$\lambda$ exchange in the ``s--channel"; one thus gets an amplitude
proportional to the imaginary part of $B^{-1}$:
\ba
S_2^{[1]}&=&S_3^{[1]}=0\,,
\\
S_1^{[1]}&=&4\pi\theta(\omega-2)\frac{-q^2M^2}{(-q^2+M^2)^2}
\cdot\frac{\sh\phi}{\phi^2+\pi^2}\,,
\ea
where $\phi$ is as in (\ref{kinphi}). Note we already anticipated 
$S_3^{[1]}=0$ by starting the sum over $r$ at 1 in (\ref{wt1on}).

In the limit of small $x$ we have
\be
S_1^{[1]}\sim\frac{2\pi}{(1-M^2/q^2)^2}\cdot\frac{1}{x\ln^2x}\,,
\,\,\,\,\,x\to0\,,\,-q^2\,\,{\rm fixed}\,,
\label{s11as}
\end{equation}
consistent with the general result (\ref{smxV}) for $l=1,2$, 
the scalar Adler function in the leading order $1/n$ being just 
$A_0(z)=(1+M^2/z)^{-2}$. Note the limit is approached
extremely slowly. E.g. denoting the asymptotic
function on the rhs of (\ref{s11as}) by $S_{\rm asympt}$ one has
$S_1^{[1]}/S_{\rm asympt}=0.245,0.441,0.638,0.907$ for 
$x=10^{-5},10^{-10},10^{-20},10^{-100}$ respectively.

One also observes the limit
\be
S_1^{[1]}\sim\frac{2\pi(1-x)}{x}\cdot\frac{1}{\ln^2(-q^2/M^2)}
\,,\,\,\,\,\,-q^2\to\infty\,\,\,x\,\,{\rm fixed}\,,
\end{equation}
and the threshold behavior: 
\be
S_1^{[1]}\sim\frac{4(-q^2/M^2)^{3/2}}{\pi(1-q^2/M^2)^2x_2}\sqrt{x_2-x}
\,,\,\,\,\,\,x\to x_2\,,\,-q^2\,\,{\rm fixed}\,.
\end{equation}
We caution that the limits $n\to\infty$ and $x\to x_2$ may not commute.

The moments (\ref{moments}) have an $1/n$ expansion of the form
\be
\Mt_{l;N}(q^2)=\sum_{r=1}^{\infty}\frac{1}{n^r}\Mt_{l;N}^{[r]}(q^2)\,,
\end{equation}
One then shows (e.g. numerically) that for $-q^2\to\infty$
\ba
\Mt_{l;1}^{[1]}(q^2)&\sim 
&\frac{2\pi}{\ln(-q^2/M^2)}\,,\,\,\,\,\,\,\,l=1,2\,,
\\
\Mt_{l;N}^{[1]}(q^2)&\sim &\frac{2\pi}{N(N-1)\ln^2(-q^2/M^2)}\,\,\,\,
{\rm for}\,\,\,N>1\,,\,\,\,l=1,2\,,
\ea
consistent with the results (\ref{Mt11}) and (\ref{spin1J}). 
The $1/\ln(-q^2)$ behavior for the $N=1$ moment comes from the 
singular behavior at $x=0$. 

So far we have only obtained the leading order for the isospin $l=1,2$ 
channels.
This is because Eq.~(\ref{wt0r}) shows that to obtain the leading order 
$1/n$ approximation for $l=0$, one has to take into account also the
amplitude $S_3^{[2]}$.
To this amplitude the only contribution comes from a diagram with
two $\lambda$--exchanges in the ``t--channel"
\be
S_3^{[2]}=\frac{-4q^2}{2\pi(-q^2+M^2)^2}{\rm Im}\int\frac{\rmd^2 k}{(2\pi)^2} 
d(k+q^{\rm E})d(k-p^{\rm E})B(k)^{-2}\,,
\end{equation}
where $d(k)$ is the Euclidean bare spin propagator 
\be
d(k)=(k^2+M^2)^{-1}\,,
\end{equation}
and $q^{\rm E}=(iq_0,q_1)$ and similarly $p^{\rm E}$ with 
$p^{{\rm E}2}=-p^2=-M^2$. Using the spectral representation of 
$B(k)^{-1}$ and the cutting rules 
\footnote{One can initially introduce a UV cutoff in the spectral
integral and remove it after invoking the cutting rules} 
(see Appendix~\ref{app1on}) we get 
\be
S_3^{[2]}=\frac{-q^2}{M^2(-q^2+M^2)^2}
\frac{\theta(\omega-2)}{\omega\sqrt{\omega^2-4}}
\frac{1}{2\pi}\left[B(k_+)^{-2}+B(k_-)^{-2}\right]\,,
\end{equation}
where
\ba
k_\pm^2&=&\frac{RM^2}{y}\left[1-\frac{y}{R}\pm
\sqrt{\frac{(1-y)(1+2x_2y^2/R)}{1-y+2y/R}}\right]\,.
\\
y&=&\frac{x}{x_2(-q^2)}\,,\,\,\,\,\,\,\,\,R=\frac{-q^2}{2M^2x_2(-q^2)}\,.
\ea
Again the small $x$ limit 
\be
S_3^{[2]}\sim\frac{2\pi}{(1-M^2/q^2)^2}\cdot\frac{1}{x\ln^2x}\,,
\,\,\,\,\,x\to0\,,\,-q^2\,\,{\rm fixed}\,,
\end{equation}
is as expected. For the large $q^2$ and threshold behaviors we have
\ba
S_3^{[2]}&\sim&\frac{2\pi}{x(1-x)}\cdot\frac{1}{\ln^2(-q^2/M^2)}
\,,\,\,\,\,\,-q^2\to\infty\,,\,x\,\,{\rm fixed}\,,
\\
S_3^{[2]}&\sim&\frac{8\pi (-q^2/M^2)^{1/2}x_2}{(1-q^2/M^2)^2}
\cdot\frac{\sinh^2\varphi}{\varphi^2}\frac{1}{\sqrt{x_2-x}}
\,,\,\,\,\,\,x\to x_2\,,\,-q^2\,\,{\rm fixed}\,,
\ea
where
\be
\varphi=2\asinh\left(\frac{\sqrt{R-1}}{2}\right)\,.
\end{equation}

For the leading isospin 0 moments we then have
\be
\Mt_{0;N}^{[1]}(q^2)=\frac{2\pi x_2^{N-1}}{R^2(1-M^2/q^2)^2}
\int_0^1\rmd z\,\frac{z^N}{S(z,q^2)}
\left[\frac{\sinh^2\theta_+}{\theta_+^2}
+\frac{\sinh^2\theta_-}{\theta_-^2}\right]\,,
\label{leading0}
\end{equation}
where
\ba
k_\pm^2&=&4M^2\sinh^2\frac{\theta_\pm}{2}\,,
\\
S(z,q^2)&=&\sqrt{(1-z)(1-z+2z/R)}\,\,.
\ea
Numerically one extracts the behavior
\be
\Mt_{0;N}^{[1]}(q^2)\sim\frac{2\pi}{\ln(-q^2/M^2)}\,\,\,{\rm for}\,\,N\ge2\,,
\end{equation}
in perfect agreement with (\ref{Mt0N}). The dominant asymptotic
piece comes from the ``$k_+$" contribution in (\ref{leading0}), 
and the dominant large $q^2$ behavior originates from the 
singularity near threshold.

\subsection{The current structure functions}

We now turn to the current structure functions for which the 
non--trivial parts are more complicated to compute in the $1/n$ 
approximation than those for the spins. We have 
\be
w^{ab;cdef}(p,q)=\sum_{r=0}^\infty 
\frac{1}{n^r}
\left[-Y^{ab;cdef}W_1^{[r]}-Y^{ba;cdef}W_2^{[r]}
+X^{ab;cdef}W_3^{[r]}\right]\,,
\end{equation}
and so for the various invariant isospin channels
\be
w_l(q^2,x)=\sum_{r=0}^\infty \frac{1}{n^r}w_l^{[r]}(q^2,x)\,,
\end{equation}
with
\ba
w_0^{[r]}&=&2W_1^{[r]}+2W_2^{[r]}+W_3^{[r+1]}\,,
\\
w_1^{[r]}&=&\phantom{-}W_1^{[r]}-W_2^{[r]}\,,
\\
w_2^{[r]}&=&-W_1^{[r]}-W_2^{[r]}\,.
\ea

At leading order one has the simple diagram with the 1--particle pole in 
the $s$--channel; this contributes to the structure functions only terms
$\propto\delta(1-x)$.

We denote the amputated two--current two--spin correlation function by 
$T^{ab;cdef}_{\mu\nu}$ with coefficients in the $1/n$ expansion
$T_{s;\mu\nu}^{[r]}$ whose imaginary parts are proportional to 
the $W_s^{[r]}$. 
There are 3 types of diagrams contributing to $T_{1;\mu\nu}^{[1]}$ 
which involve one $\lambda$ propagator.
One is the box diagram, another involves a vertex correction,
and the last involves a spin self energy diagram. 
They can be conveniently combined together
\footnote{one can conveniently use lattice UV regularization at
intermediate stages}
to yield (in Euclidean space)
\be
T_{1;\mu\nu}^{[1]}(p^{\rm E},q^{\rm E})=8\int\frac{\rmd^2 k}{(2\pi)^2}
D_\mu(p^{\rm E},q^{\rm E},k)D_\nu(p^{\rm E},q^{\rm E},k)
d(k+q^{\rm E})B(p^{\rm E}-k)^{-1}\,, 
\end{equation} with
\be
D_\mu(p,q,k)=(k+q/2)_\mu d(k)+(p+q/2)_\mu d(p+q)\,.
\end{equation}
So contracting over $\mu,\nu$ 
\ba
&&\sum_\mu T_{1;\mu\mu}^{[1]}(p,q)=F(p)
+8\int\frac{\rmd^2 k}{(2\pi)^2}D(p,q,k)d(k+q)B(p-k)^{-1}
\nonumber\\
&&
\\
&&D(p,q,k)=(p+q/2)^2d(p+q)^2+\frac12[d(p+q)+d(k)]
\nonumber\\
&&+[-(p-k)^2+p^2+pq-M^2]d(p+q)d(k)-\frac14[q^2+4M^2]d(k)^2\,,
\ea
where $F$ is a function of $p^2$ alone.
Since we are only interested in the structure function only one
term appearing in the cutting rule is relevant. Still in Euclidean space,
and omitting pole terms $\propto\delta(1-x)$, the relevant term is
\be
\sum_\mu T_{1;\mu\mu}^{[1]}(p^{\rm E},q^{\rm E})\sim
-16\pi M^4\int_0^\Lambda\rmd\kappa 
\frac{\sinh^2\kappa}{\kappa^2+\pi^2}
B(p^{\rm E}+q^{\rm E};M,m)\left(I_++I_-\right)\,,
\end{equation}
where $\Lambda$ is some ultraviolet cutoff and
\ba
m&=&2M\cosh\kappa/2\,,
\\
I_\pm&=&2D(p^{\rm E},q^{\rm E},k_\pm(p^{\rm E},q^{\rm E}))\,, 
\\
k_\pm^2(p,q)&=&-qp-\frac{[M^2(p^2+qp)+m^2(q^2+qp)]}{(q+p)^2}
\nonumber\\
&&\mp\frac{i\epsilon qp}{(q+p)^2}
\sqrt{\left[(p+q)^2+(m+M)^2\right]\left[(p+q)^2+(m-M)^2\right]}\,.
\ea

So going over to Minkowski space
\be
W_1^{[1]}=4M^4\theta(W-3M)\int_0^{\kappa_0}\rmd\kappa
\frac{(C_++C_-)\sinh^2\kappa}
{(\kappa^2+\pi^2)\sqrt{\left[W^2-(m+M)^2\right]\left[W^2-(m-M)^2\right]}}
\end{equation}
with $\kappa_0$ defined through
\be
\cosh\frac{\kappa_0}{2}=\frac{W-M}{2M}
\end{equation}
and
\ba
C_\pm&=&-\frac{4M^2+4pq+q^2}{2(W^2-M^2)^2}
-\frac{1}{W^2-M^2}\left[1+2(-pq+m^2-2M^2)d(K_\pm)\right]
\nonumber\\
&&+d(K_\pm)-\frac12(-q^2+4M^2)d(K_\pm)^2\,,
\\
K_\pm^2&=&qp-\frac{[M^2(M^2+qp)+m^2(q^2+qp)]}{W^2}
\nonumber\\   
&&\mp\frac{\sqrt{(pq)^2-M^2q^2}}{W^2}
\sqrt{\left[W^2-(m+M)^2\right]\left[W^2-(m-M)^2\right]}\,.
\ea

Numerically for small $x$ we find consistency 
\footnote{we did not yet confirm this analytically}
with the general result (\ref{smxT}):
\be
w_l^{[1]}\sim\frac{2\pi}{x\ln^2x}a_1(-q^2)\,,\,\,\,\,\,l=1,2\,,
\end{equation} 
with
\ba
a_1(-q^2)&=&\frac{1}{2\pi}\left[3-
\frac{\theta}{\sinh\theta}\left(2+\cosh\theta\right)\right]\,,
\\
-q^2&=&4M^2\sinh^2\frac{\theta}{2}\,,
\ea
since $a_1(z)$ is the leading order $1/n$ contribution to the 
Adler function $A_1(z)$. In leading order $1/n$ the current vacuum
2--point function amplitude $I_1=i_1$ with
\footnote{For $z=-q^2$ we have 
$-z^2\frac{\partial}{\partial z} i_1(z)=\frac{1}{\pi}+
\frac{z^2}{2M^4\sinh\theta}\frac{\partial}{\partial\theta}
\left[\frac{\theta\cosh\frac{\theta}{2}}{8\pi\sinh^3\frac{\theta}{2}}
\right]=a_1(z)$}
\be
i_1(z)=\frac{1}{\pi z}-\frac{(z+4M^2)}{z}B(r)\,,\,\,\,\,r^2=z\,.
\end{equation}

We have not yet computed the $1/n$ contribution to the isospin
0 structure function $w_0$. This also involves computing 
$T_{3\mu\nu}^{[2]}$ 
which is more complicated because it requires the evaluation of 2--loop 
graphs involving also $\lambda$ propagators.

\section{Summary and Conclusions}

In this paper we calculated the DIS structure functions (and their
moments) in the family of the 2--dimensional O$(n)$ nonlinear sigma
models using standard field theory techniques available in any
asymptotically free field theory model. In the special case of the 
O$(3)$ model we compared the results to the non--perturbative (bootstrap)
determination of the same structure functions.  

The very good agreement between the results (in the intermediate 
energy range, where both the perturbative field theory results and
the non--perturbative bootstrap results are expected to be valid)
strongly indicates -- once again -- that standard field theory and the 
bootstrap define the same model. 
On the other hand this agreement provides some
further, indirect proof for all the assumptions that are used in the
derivations in both methods. 

The study of the structure functions has lead us to two interesting
findings. First, we found that the isospin 0 structure functions
exhibit exact Bjorken scaling: for $-q^2\rightarrow\infty$ 
the structure functions go to some non--trivially $x$--dependent limits. 
Second, the exact small $x$ asymptotics of the structure functions are
shown to be different from the soft--Pomeron like fractional power
behavior: the asymptotics of ($x$ times) the structure function is
logarithmic. 

In the first case we have obtained concrete results in the O$(3)$
model only but we think that our findings are more generally valid: 
it is probable that exact Bjorken scaling is due to the presence of the 
(infinitely many) higher spin conserved charges characteristic  
to integrable models. Also in the second case we believe that  
the small $x$ asymptotics we found here is valid in a more general setting. 
Whether something similar applies to QCD remains to be seen.

\section*{Acknowledgements}

J. B. is grateful to the Max-Planck-Institut f\"ur Physik for its
hospitality.
This investigation was supported in part by the Hungarian 
National Science Fund OTKA (under T034299 and T043159).

\appendix

\section{$\On$ notations and identities}

The $\On$ generators $Q^{ab}$ act in the defining (vector)
representation as
\be
\left[Q^{ab},V^c\right]=it^{ab}_{cd}\,V^d\,,
\end{equation}
where the generator matrix is
\be
t^{ab}_{cd}=\delta^{ac}\delta^{bd}-\delta^{ad}\delta^{bc}\,.
\label{tdef}
\end{equation}
This corresponds to the usual relation
\be
\left[Q^{a},V^b\right]=i\epsilon^{abc}\,V^c
\end{equation}
in the $n=3$ case with $Q^{a}=\frac{1}{2}\epsilon^{abc}\,Q^{bc}$.

The generator matrix in the 2--index tensor representation is
\be
t^{ab}_{cd;uv}=t^{ab}_{cu}\,\delta^{dv}+\delta^{cu}\,t^{ab}_{dv}\,,
\end{equation}
and similarly for higher representations.

Projector matrices in the 2--index tensor representation are
\ba
P_0^{ab;a^\prime b^\prime}&=& 
\frac{1}{n}\delta^{ab}\delta^{a^\prime b^\prime},
\label{pdef0}
\\
P_1^{ab;a^\prime b^\prime}&=& 
\frac{1}{2}\left(\delta^{aa^\prime}\delta^{bb^\prime}-
\delta^{ab^\prime}\delta^{ba^\prime}\right),\\
P_2^{ab;a^\prime b^\prime}&=& 
\frac{1}{2}\left(\delta^{aa^\prime}\delta^{bb^\prime}+
\delta^{ab^\prime}\delta^{ba^\prime}\right)
-\frac{1}{n}\delta^{ab}\delta^{a^\prime b^\prime}\,.
\label{pdef2}
\ea
They satisfy
\ba
P_k^{ab;a^\prime b^\prime}
P_l^{a^\prime b^\prime;a^{\prime\prime} b^{\prime\prime}}
&=&\delta_{kl}\,P_l^{ab;a^{\prime\prime} b^{\prime\prime}}\,,
\\
\sum_l P_l^{ab;a^\prime b^\prime}&=&
\delta^{aa^\prime}\delta^{bb^\prime}\,,
\ea
and
\be
P_k^{ab;ab}=\pi_l
\end{equation}
with
\be
\pi_0=1,\qquad\qquad
\pi_1=\frac{n(n-1)}{2},\qquad\qquad
\pi_2=\frac{(n-1)(n+2)}{2}\,.
\end{equation}
In the $n=3$ case $\pi_l=2l+1$.

6--index invariant tensors, antisymmetric in the last two index pairs are
\be
X^{ab;cdef}=\delta^{ab}\left(\delta^{ce}\delta^{df}-
\delta^{cf}\delta^{de}\right)\,,
\end{equation}
\be
Y^{ab;cdef}=\delta^{ac}\delta^{be}\delta^{df}-
\delta^{ad}\delta^{be}\delta^{cf}-
\delta^{ac}\delta^{bf}\delta^{de}+
\delta^{ad}\delta^{bf}\delta^{ce}\,,
\end{equation}
(and $Y^{ba;cdef})$. The irreducible combinations are
\ba
R_0^{ab;cdef}&=&\frac{1}{n}X^{ab;cdef},\\ 
R_1^{ab;cdef}&=&\frac{1}{2}\left(Y^{ab;cdef}-Y^{ba;cdef}\right),\\ 
R_2^{ab;cdef}&=&\frac{2}{n}X^{ab;cdef}-
\frac{1}{2}\left(Y^{ab;cdef}+Y^{ba;cdef}\right)\,, 
\ea
which satisfy
\be
R_k^{ab;cdef}
R_l^{a^\prime b^\prime;cdef}
=\delta_{kl}\,\hat r_lP_l^{ab;a^\prime b^\prime}
\end{equation}
with
\be
\hat r_0=2(n-1)\,,\qquad\qquad
\hat r_1=\hat r_2=4(n-2)\,.
\end{equation}
Note that in the $n=3$ case the antisymmetric tensor representation 
coincides with the vector representation and indeed in this case
\be
\frac{1}{4}\epsilon^{cxy}\epsilon^{duv}R_l^{ab;xyuv}=
P_l^{ab;cd}\,.
\end{equation}

Finally we note the following identities. The generators of the vector
representation satisfy
\be
t^{ax}_{cu}\,t^{bx}_{du}=\sum_{l=0}^2 V_l\,P_l^{ab;cd}
\label{Vdef}
\end{equation}
with
\be
V_0=2(n-1)\,,\qquad\qquad V_1=V_2=n-2\,,
\label{vl}
\end{equation}
and similarly for the antisymmetric tensor representation
\be
t^{ax}_{cd;uv}\,t^{bx}_{ef;uv}-t^{ax}_{cd;uv}\,t^{bx}_{ef;vu}
=\sum_{l=0}^2 T_l\,R_l^{ab;cdef}\,,
\label{Tdef}
\end{equation}
where
\be
T_0=4(n-2)\,,\qquad\qquad T_1=n-2\,,\qquad\qquad T_2=4-n\,.
\label{tl}
\end{equation}

\section{Particle states, rapidity integrals}

The $r$--particle \lq\lq in" states are characterized by the $\On$
labels $a_1,\dots,a_r$ and the decreasing set of rapidities 
$\theta_1,\dots,\theta_r$ and their normalization is
\be
\begin{split}
&^{\rm in}\langle
a_1^\prime,\theta_1^\prime;\dots;a_r^\prime,\theta_r^\prime
\vert a_1,\theta_1;\dots;a_r,\theta_r\rangle^{\rm in}=\\
&\qquad\qquad\qquad
\qquad\qquad\qquad
(4\pi)^r\delta^{a_1a_1^\prime}\dots\delta^{a_ra_r^\prime}
\delta(\theta_1^\prime-\theta_1)\dots \delta(\theta_r^\prime-\theta_r)\,,
\end{split}
\end{equation}
corresponding to the completeness relation in the $r$--particle sector
\be
\begin{split}
&\prod{}^{(r)}=\frac{1}{(4\pi)^r}\sum_{a_1\dots a_r}
\int_{-\infty}^{\infty}\rmd\theta_1
\int_{-\infty}^{\theta_1}\rmd\theta_2\dots\\
&\qquad\qquad\qquad\qquad\qquad
\dots\int_{-\infty}^{\theta_{r-1}}\rmd\theta_r\,
\vert a_1,\theta_1;\dots;a_r,\theta_r\rangle^{\rm in}\,
^{\rm in}\langle a_1,\theta_1;\dots;a_r,\theta_r\vert\,.
\end{split}
\end{equation}

As usual, we introduce the set of positive rapidity differences
\be
u_1=\theta_1-\theta_2\,,\qquad
u_2=\theta_2-\theta_3\,,\qquad\dots\qquad
u_{r-1}=\theta_{r-1}-\theta_r\qquad
\end{equation}
and the $r$--particle invariant mass $M^{(r)}(u)$ with the definition
\be
M^{(r)}(u)\rme^{\pm\Lambda}=M\sum_{i=1}^r\rme^{\pm\theta_i}=E_r\pm P_r
=P^0_r\pm P^1_r\,,
\end{equation}
where $M$ is the mass of the $\On$ particles.

The $r$--particle rapidity integral can now be written
\be
\int_{-\infty}^{\infty}\rmd\theta_1
\int_{-\infty}^{\theta_1}\rmd\theta_2\,\,\dots
\int_{-\infty}^{\theta_{r-1}}\rmd\theta_r=
\int_{-\infty}^\infty\rmd\Lambda\int\,{\cal D}u^{(r)}\,,
\end{equation}
where
\be
{\cal D}u^{(r)}=
\int_0^\infty\rmd u_1 
\int_0^\infty\rmd u_2\,\,\dots 
\int_0^\infty\rmd u_{r-1}\,. 
\end{equation}

The inverse transformation is
\be
\theta_i=\beta_i+\Lambda-v^{(r)}_++v^{(r)}_-\,,
\end{equation}
where
\ba
\beta_j&=&u_{jr}\,,\,\,\,\,j=1,\dots,r-1\,,
\\
u_{jk}&=&u_j+u_{j+1}+\dots+u_{k-1}\,,\,\,\,\,\,\,1\le j<k\le r\,,
\\
\beta_r&=&0\,,
\ea
and
\be
v^{(r)}_\pm=\frac{1}{2}\ln\left[1+\sum_{i=1}^{r-1}\,\rme^{\pm\beta_i}\right]\,.
\end{equation}
We note that
\be
v^{(r)}_++v^{(r)}_-=\ln\mu_r\,,
\end{equation}
where $\mu_r=M^{(r)}(u)/M$ is the dimensionless invariant mass.

\section{S--matrix asymptotics}

The Zamolodchikov $\On$ S--matrix is \cite{ZZ} 
\be
S_{ab;cd}(\theta)=\sigma_1(\theta)\delta^{ab}\delta^{cd}+
\sigma_2(\theta)\delta^{ac}\delta^{bd}+
\sigma_3(\theta)\delta^{ad}\delta^{bc}\,,
\end{equation}
where
\ba
\sigma_1(\theta)&=&
\frac{-2\pi i\theta}{i\pi-\theta}\,
\frac{s_2(\theta)}{(n-2)\theta-2\pi i}\,,\\
\sigma_2(\theta)&=&
(n-2)\theta\,
\frac{s_2(\theta)}{(n-2)\theta-2\pi i}\,,\\
\sigma_3(\theta)&=&
-2\pi i\,
\frac{s_2(\theta)}{(n-2)\theta-2\pi i}\,,
\ea
and the \lq\lq isospin 2" phase shift is given by
\be
s_2(\theta)=-\exp\left\{2i\int_0^\infty\,\frac{\rmd\omega}{\omega}\,
\sin(\theta\omega)\,\tilde K_n(\omega)\right\}
\end{equation}
with
\be
\tilde K_n(\omega)=\frac{\rme^{-\pi\omega}+\rme^{-2\pi\frac{\omega}{n-2}}}
{1+\rme^{-\pi\omega}}\,.
\end{equation}
Specially for $n=3$
\be
s_2(\theta)=\frac{\theta-i\pi}{\theta+i\pi}\,.
\end{equation}

Using the asymptotic formula
\be
\int_0^\infty\,\frac{\rmd\omega}{\omega}\,
\sin(\theta\omega)\,k(\omega) \cong
\frac{\pi}{2}\,k(0)+\frac{k^\prime(0)}{\theta}+
\rmO\left(\frac{1}{\theta^2}\right)\,,
\end{equation}
we get the large $\theta$ asymptotics of the S--matrix, which can be
written as
\be
S_{ab;cd}(\theta)\cong \delta^{ac}\delta^{bd}+
\frac{2\pi i}{(n-2)\theta}\,t^{ac}_{bd}
+\rmO\left(\frac{1}{\theta^2}\right)\,.
\label{asyS}
\end{equation}

\section{Residue asymptotics}

For any $(r+2)$--particle form factor in the $\On$ model
Smirnov's residue axiom \cite{Smirnov} can be written as
\be
\begin{split}
&{\cal F}^A_{aba_1\dots a_r}(\beta+i\pi+\varepsilon,\beta,\theta_1,\dots,
\theta_r)\cong\frac{2i}{\varepsilon}\Big\{
\delta_{ab}\,{\cal F}^A_{a_1\dots a_r}(\theta_1,\dots,\theta_r)\\
&\qquad\qquad
-S_{ba_1\dots a_r;b_1\dots b_ra}(\theta_1,\dots,\theta_r\vert\beta)
\,{\cal F}^A_{b_1\dots b_r}(\theta_1,\dots,\theta_r)\Big\}\,,
\end{split}
\end{equation}
where
\be
\begin{split}
&S_{ba_1\dots a_r;b_1\dots b_ra}(\theta_1,\dots,\theta_r\vert\beta)=
S_{ba_1;c_1b_1}(\beta-\theta_1)
S_{c_1a_2;c_2b_2}(\beta-\theta_2)\cdots\\
&\qquad\qquad\cdots
S_{c_{r-2}a_{r-1};c_{r-1}b_{r-1}}(\beta-\theta_{r-1})
S_{c_{r-1}a_r;ab_r}(\beta-\theta_r)\,.
\end{split}
\end{equation}
If $\beta$ is large we can use (\ref{asyS}) to get
\be
\begin{split}
&{\cal F}^A_{aba_1\dots a_r}(\beta+i\pi+\varepsilon,\beta,
\theta_1,\dots,\theta_r)\cong\\ 
&\qquad\qquad
\qquad\qquad
\qquad
-\frac{4\pi}{(n-2)\varepsilon\beta}\,\sum_{i=1}^r\, t^{ab}_{a_ib_i}
\,{\cal F}^A_{a_1\dots b_i\dots a_r}(\theta_1,\dots,\theta_r)\,.
\end{split}
\label{asyRES1}
\end{equation}

So far the operator index $A$ did not play any role. For the case of
tensor operators, where $A$ is an $\On$ (multi) index, the form
factors are invariant tensors and (\ref{asyRES1}) can equivalently
be written
\be
{\cal F}^A_{aba_1\dots a_r}(\beta+i\pi+\varepsilon,\beta,
\theta_1,\dots,\theta_r)\cong 
\frac{4\pi}{(n-2)\varepsilon\beta}\, t^{ab}_{AB}
\,{\cal F}^B_{a_1\dots a_r}(\theta_1,\dots,\theta_r)\,,
\label{asyRES2}
\end{equation}
where $t^{ab}_{AB}$ is the $\On$ generator in the appropriate 
representation.

\section{Operator basis}
\label{appE}

In the operator product expansions we use a basis spanned by
${\cal A}^{(l)ab}_\omega$,
${\cal B}^{(J,l)ab}_\omega$ and
$\overline{{\cal B}}^{(J,l)ab}_\omega$, where these basis elements are
hermitian local operators at $(0,0)$ and $l=0,1$ or $2$ tensor operators
(in the index pair $ab$) under $\On$.
Under the action of the parity operator $V$ they transform as
\be
V{\cal A}^{(l)ab}_\omega V={\cal A}^{(l)ab}_\omega\,,\qquad\qquad
V{\cal B}^{(J,l)ab}_\omega V=
\overline{{\cal B}}^{(J,l)ab}_\omega.
\end{equation}
Their Lorentz spin can be read off the relations
\be
\begin{align}
\left[{\cal M},{\cal A}^{(l)ab}_\omega\right]
&=0\,,\\
\left[{\cal M},{\cal B}^{(J,l)ab}_\omega\right]
&=\phantom{-}iJ{\cal B}^{(J,l)ab}_\omega\,,\\
\left[{\cal M},\overline{{\cal B}}^{(J,l)ab}_\omega\right]
&=-iJ\overline{{\cal B}}^{(J,l)ab}_\omega\,,
\end{align}
\end{equation}
where $J$ is a positive integer and ${\cal M}$ is the Lorentz boost 
operator.
Finally under the action of the (anti-linear) CPT operator $\Theta$
\be
\begin{align}
\Theta{\cal A}^{(l)ab}_\omega\Theta &=(-1)^l\,
{\cal A}^{(l)ab}_\omega,\\
\Theta{\cal B}^{(J,l)ab}_\omega\Theta &=(-1)^l\,
{\cal B}^{(J,l)ab}_\omega,\\
\Theta\overline{{\cal B}}^{(J,l)ab}_\omega\Theta &=(-1)^l\,
\overline{{\cal B}}^{(J,l)ab}_\omega.
\end{align}
\end{equation}
The one-particle matrix elements of the above operators are
parametrized as
\be
\begin{align}
\langle a,\theta\vert{\cal A}^{(l)cd}_\omega\vert b,\theta\rangle&=
P^{ab;cd}_l\,A^{(l)}_\omega,\\
\langle a,\theta\vert{\cal B}^{(J,l)cd}_\omega\vert b,\theta\rangle&=
\left(-\frac{iM}{2}\rme^\theta\right)^J\,P^{ab;cd}_l\,
B^{(J,l)}_\omega,\\
\langle a,\theta\vert
\overline{{\cal B}}^{(J,l)cd}_\omega\vert b,\theta\rangle&=
\left(-\frac{iM}{2}\rme^{-\theta}\right)^J\,P^{ab;cd}_l\,
B^{(J,l)}_\omega\,.
\end{align}
\end{equation}
Note that we have considered operators with non--vanishing
one--particle matrix elements only.

\section{Notations and Conventions}

We will use the notation
\be
W_2=-iW_0
\label{W20}
\end{equation}
for any vector (and higher tensor) index. The light--cone components are
\be
W_\pm=\frac{1}{2}\left(W_0\mp W_1\right)
=\frac{1}{2}\left(iW_2\mp W_1\right)
\label{LC}
\end{equation}
and similarly
\be
\partial_\pm=\frac{1}{2}\left(\partial_0\mp \partial_1\right)
=\frac{1}{2}\left(i\partial_2\mp \partial_1\right).
\end{equation}
We treat the two--dimensional Euclidean coordinates exceptionally since 
here we use
\be
y_\pm=\mp\, y_1-iy_2,
\label{ycoord}
\end{equation}
which gives
\be
y_+\,y_-=-y^2,
\end{equation}
as opposed to the Euclidean square of vectors in (\ref{LC}), which is given by
\be
\left(W^2\right)_{\rm E}=W_1^2+W_2^2=-4W_+\,W_-.
\label{square}
\end{equation}

Two--dimensional Fourier transformation is indicated by tilde:
\be
\tilde f(Q)=\int\rmd^2y\,\,\rme^{iQy}\,f(y)\,.
\label{Fourier}
\end{equation}
For functions depending on $y^2$ only we also define
\be
\hat f(Q)=\left(Q^2\right)^{J+1}\left(\frac{\rmd}{\rmd Q^2}\right)^J
\int\rmd^2y\,\,\rme^{ipy}\,f(y^2\,).
\label{fhat}
\end{equation}

\section{Asymptotic expansions}
\label{Asy}

Assume that $S(y)$ has an asymptotic expansion
\be
S(y)=f_0\,\lambdat^{\sigma-1}+\rmO\left(\lambdat^\sigma\right)
\end{equation}
in terms of the effective coupling $\lambdat$ defined in (\ref{lambdat}).
Then $\hat{S}(Q)$ can be asymptotically expanded as
\be
\hat{S}(Q)=(2\pi)(-1)^JJ!(1-\sigma)f_0\,\lambda^\sigma
+\rmO\left(\lambda^{\sigma+1}\right)
\end{equation}
in terms of the effective coupling $\lambda$ defined by
\be
\frac{1}{\lambda}+\frac{1}{n-2}\,\ln\lambda=\ln\frac{|Q|}{
\Lambda_{\overline{\rm MS}}}\,.
\end{equation}
In the special case
\be
S(y)=f_0\,\frac{1}{\lambdat}+\rmO(1)\,,
\end{equation}
we have
\be
\hat S(Q)=(2\pi)(-1)^JJ!\, f_0\,(1+\frac{1}{n-2}\lambda)+\rmO(\lambda^2)
\end{equation}
i.e. in this case we also know the coefficient of the sub-leading term.
Finally if
\be
S(y)=f_0\,\ln\lambdat+\rmO(\lambdat)
\end{equation}
then
\be
\hat S(Q)=-(2\pi)(-1)^JJ!\,f_0\,\lambda+\rmO(\lambda^2)\,.
\end{equation}

An alternative way of presenting the above results is as follows.  
If the derivative of the function $W(y^2)$ has the asymptotic expansion
\be
y^2\frac{\rmd}{\rmd y^2}W(y^2)=
\alpha\lambdat^\sigma+\rmO\left(\lambdat^{\sigma+1}\right)
\end{equation}
then in Fourier space we have
\be
\hat W(Q^2)=-4\pi\alpha(-1)^J J!\lambda^\sigma
+\rmO\left(\lambda^{\sigma+1}\right)\,.
\end{equation}
In the special case $\sigma=0$ if  
\be
y^2\frac{\rmd}{\rmd y^2} W(y^2)=\alpha+\beta\lambdat
+\rmO\left(\lambdat^2\right)
\end{equation}
then
\be
\hat W(Q^2)=-4\pi(-1)^J J!\left\{\alpha+\beta\lambda+
\rmO\left(\lambda^2\right)\right\}\,.
\end{equation}

\section{Reduced spin and current form factor squares}
\label{apph}

The space of homogeneous symmetric polynomials in $r$ variables
$\theta_i\,,i=1\dots r$ is spanned by products of 
$\sigma_k^{(r)}\,,\,\,1\le k\le r$
\be
\sigma_k^{(r)}=\sum_{1\le i_1<\dots <i_k\le r}
\theta_{i_1}\dots\theta_{i_k}\,.
\label{polygens}
\end{equation}

For the reduced spin and current structure functions we have 

for $r=2$:

\ba
j_0^{(2)}&=&4\left(\sigma_1^2-3\sigma_2\right)+8\pi^2\,,
\label{j0}\\
j_1^{(2)}&=&\sigma_1^2-2\sigma_2-2\pi^2\,,
\\
j_2^{(2)}&=&\sigma_1^2-6\sigma_2+2\pi^2\,.
\label{j2}
\ea

for $r=3$:

\ba
j_0^{(3)}&=&4\left(-6\sigma_1^3\sigma_3+2\sigma_1^2\sigma_2^2
+19\sigma_1\sigma_2\sigma_3-6\sigma_2^3-9\sigma_3^2\right)
\nonumber
\\
&&+4\pi^2\left(4\sigma_1^4-21\sigma_1^2\sigma_2
-19\sigma_1\sigma_3+34\sigma_2^2\right)
\nonumber
\\
&&+8\pi^4\left(9\sigma_1^2-22\sigma_2\right)+64\pi^6\,,
\\
j_1^{(3)}&=&-2\sigma_1^3\sigma_3+\sigma_1^2\sigma_2^2
+2\sigma_1\sigma_2\sigma_3-2\sigma_2^3+9\sigma_3^2
\nonumber
\\
&&-2\pi^2\left(\sigma_1^4-3\sigma_1^2\sigma_2
+10\sigma_1\sigma_3-4\sigma_2^2\right)
\nonumber
\\
&&-2\pi^4\left(3\sigma_1^2-\sigma_2\right)-8\pi^6\,,
\\ 
j_2^{(3)}&=&6\sigma_1^3\sigma_3-\sigma_1^2\sigma_2^2
-38\sigma_1\sigma_2\sigma_3+6\sigma_2^3+99\sigma_3^2       
\nonumber
\\
&&-2\pi^2\left(\sigma_1^4-9\sigma_1^2\sigma_2
+8\sigma_1\sigma_3+16\sigma_2^2\right)
\nonumber
\\
&&-2\pi^4\left(3\sigma_1^2-17\sigma_2\right)-8\pi^6\,.
\label{j3}
\ea

for $r=4$:

\ba
j_0^{(4)}&=&16\sigma_1^2\sigma_2^2\sigma_3^2-48\sigma_2^3\sigma_3^2
-48\sigma_1^3\sigma_3^3+152\sigma_1\sigma_2\sigma_3^3-72\sigma_3^4
\nonumber
\\ 
&&-(48\sigma_1^2\sigma_2^3-144\sigma_2^4-152\sigma_1^3\sigma_2\sigma_3
+476\sigma_1\sigma_2^2\sigma_3-56\sigma_1^2\sigma_3^2
-52\sigma_2\sigma_3^2)\sigma_4
\nonumber
\\
&&-(72\sigma_1^4-52\sigma_1^2\sigma_2
-352\sigma_2^2-128\sigma_1\sigma_3)\sigma_4^2-640\sigma_4^3
\nonumber
\\
&&+4\pi^2[8\sigma_1^2\sigma_2^4-24\sigma_2^5
-42\sigma_1^3\sigma_2^2\sigma_3+131\sigma_1\sigma_2^3\sigma_3
+68\sigma_1^4\sigma_3^2
\nonumber
\\ 
&&-249\sigma_1^2\sigma_2\sigma_3^2
+73\sigma_2^2\sigma_3^2+55\sigma_1\sigma_3^3
-(38\sigma_1^4\sigma_2-279\sigma_1^2\sigma_2^2+528\sigma_2^3
\nonumber
\\
&&+175\sigma_1^3\sigma_3
-669\sigma_1\sigma_2\sigma_3+133\sigma_3^2)\sigma_4
-(157\sigma_1^2+32\sigma_2)\sigma_4^2]
\nonumber 
\\ 
&&+4\pi^4[36\sigma_1^4\sigma_2^2-184\sigma_1^2\sigma_2^3
+260\sigma_2^4-88\sigma_1^5\sigma_3+447\sigma_1^3\sigma_2\sigma_3
\nonumber
\\ 
&&+253\sigma_1^2\sigma_3^2-696\sigma_1\sigma_2^2\sigma_3
-248\sigma_2\sigma_3^2
\nonumber
\\
&&+(185\sigma_1^4-1156\sigma_1^2\sigma_2+2120\sigma_2^2
-740\sigma_1\sigma_3)\sigma_4+96\sigma_4^2]
\nonumber
\\
&&+4\pi^6[32\sigma_1^6-232\sigma_1^4\sigma_2+780\sigma_1^2\sigma_2^2
-992\sigma_2^3-532\sigma_1^3\sigma_3
+1269\sigma_1\sigma_2\sigma_3
\nonumber
\\ 
&&+295\sigma_3^2
+(1211\sigma_1^2-3220\sigma_2)\sigma_4]
\nonumber
\\
&&+16\pi^8[60\sigma_1^4-278\sigma_1^2\sigma_2+405\sigma_2^2
-212\sigma_1\sigma_3+434\sigma_4]
\nonumber
\\
&&+128\pi^{10}[17\sigma_1^2-37\sigma_2]+1280\pi^{12}\,,
\label{j40}
\ea
\ba
j_1^{(4)}&=&\sigma_1^2\sigma_2^2\sigma_3^2-2\sigma_2^3\sigma_3^2
-2\sigma_1^3\sigma_3^3+2\sigma_1\sigma_2\sigma_3^3+9\sigma_3^4
\nonumber
\\
&&+(-2\sigma_1^2\sigma_2^3+4\sigma_2^4+2\sigma_1^3\sigma_2\sigma_3
+2\sigma_1\sigma_2^2\sigma_3+18\sigma_1^2\sigma_3^2
-62\sigma_2\sigma_3^2)\sigma_4
\nonumber
\\
&&+(9\sigma_1^4-62\sigma_1^2\sigma_2+80\sigma_2^2)\sigma_4^2+128\sigma_4^3
\nonumber
\\ 
&&-2\pi^2[\sigma_1^2\sigma_2^4-2\sigma_2^5-3\sigma_1^3\sigma_2^2\sigma_3
+3\sigma_1\sigma_2^3\sigma_3-4\sigma_1^4\sigma_3^2
+33\sigma_1^2\sigma_2\sigma_3^2-39\sigma_2^2\sigma_3^2
\nonumber
\\
&&+7\sigma_1\sigma_3^3
+(10\sigma_1^4\sigma_2-45\sigma_1^2\sigma_2^2+78\sigma_2^3
-15\sigma_1^3\sigma_3-36\sigma_1\sigma_2\sigma_3-118\sigma_3^2)\sigma_4
\nonumber
\\
&&+(-14\sigma_1^2+264\sigma_2)\sigma_4^2]
\nonumber
\\
&&+2\pi^4[-3\sigma_1^4\sigma_2^2+12\sigma_1^2\sigma_2^3-16\sigma_2^4
+\sigma_1^5\sigma_3+\sigma_1^3\sigma_2\sigma_3+58\sigma_1^2\sigma_3^2
\nonumber
\\
&&
+9\sigma_1\sigma_2^2\sigma_3-174\sigma_2\sigma_3^2
+(15\sigma_1^4-202\sigma_1^2\sigma_2+456\sigma_2^2-256\sigma_1\sigma_3)\sigma_4
+1184\sigma_4^2]
\nonumber
\\
&&-2\pi^6[4\sigma_1^6-24\sigma_1^4\sigma_2+57\sigma_1^2\sigma_2^2
-26\sigma_2^3+37\sigma_1^3\sigma_3
\nonumber   
\\
&&-198\sigma_1\sigma_2\sigma_3+26\sigma_3^2
+(-272\sigma_1^2+1100\sigma_2)\sigma_4]
\nonumber
\\
&&+8\pi^8[-3\sigma_1^4-2\sigma_1^2\sigma_2+13\sigma_2^2+3\sigma_1\sigma_3
+126\sigma_4]
\nonumber
\\
&&-256\pi^{10}\sigma_2+128\pi^{12}\,,  
\label{j41}
\ea
\ba
j_2^{(4)}&=&
\sigma_1^2\sigma_2^2\sigma_3^2-6\sigma_2^3\sigma_3^2-6\sigma_1^3\sigma_3^3
+38\sigma_1\sigma_2\sigma_3^3-99\sigma_3^4
\nonumber
\\
&&+(-6\sigma_1^2\sigma_2^3+36\sigma_2^4+38\sigma_1^3\sigma_2\sigma_3
-242\sigma_1\sigma_2^2\sigma_3+2\sigma_1^2\sigma_3^2
+670\sigma_2\sigma_3^2)\sigma_4
\nonumber
\\
&&+(-99\sigma_1^4+670\sigma_1^2\sigma_2-656\sigma_2^2
-1600\sigma_1\sigma_3)\sigma_4^2+3584\sigma_4^3
\nonumber
\\   
&&-2\pi^2[-\sigma_1^2\sigma_2^4+6\sigma_2^5+9\sigma_1^3\sigma_2^2\sigma_3
-55\sigma_1\sigma_2^3\sigma_3-16\sigma_1^4\sigma_3^2
\nonumber
\\
&&+99\sigma_1^2\sigma_2\sigma_3^2
+103\sigma_2^2\sigma_3^2-257\sigma_1\sigma_3^3
\nonumber
\\
&&+(-8\sigma_1^4\sigma_2+21\sigma_1^2\sigma_2^2+78\sigma_2^3
+125\sigma_1^3\sigma_3-762\sigma_1\sigma_2\sigma_3+1664\sigma_3^2)\sigma_4
\nonumber
\\
&&+(248\sigma_1^2-824\sigma_2)\sigma_4^2]
\nonumber
\\
&&+2\pi^4[3\sigma_1^4\sigma_2^2-32\sigma_1^2\sigma_2^3
+64\sigma_2^4-17\sigma_1^5\sigma_3+165\sigma_1^3\sigma_2\sigma_3
\nonumber
\\
&&-280\sigma_1^2\sigma_3^2
-303\sigma_1\sigma_2^2\sigma_3+632\sigma_2\sigma_3^2
\nonumber
\\
&&+(55\sigma_1^4-284\sigma_1^2\sigma_2+160\sigma_2^2
+632\sigma_1\sigma_3)\sigma_4-960\sigma_4^2]
\nonumber
\\
&&-2\pi^6[-4\sigma_1^6+56\sigma_1^4\sigma_2-201\sigma_1^2\sigma_2^2
+238\sigma_2^3-37\sigma_1^3\sigma_3
\nonumber
\\
&&-36\sigma_1\sigma_2\sigma_3+472\sigma_3^2+(-82\sigma_1^2
+260\sigma_2)\sigma_4]
\nonumber
\\
&&+8\pi^8[3\sigma_1^4-46\sigma_1^2\sigma_2+93\sigma_2^2
+35\sigma_1\sigma_3+58\sigma_4]
\nonumber
\\
&&+64\pi^{10}[\sigma_1^2-8\sigma_2]+128\pi^{12}\,.
\nonumber
\\
&&
\label{j42}
\ea

\hfill
\eject

\section{Structure function moments}
\label{momtables}

\vspace{20mm}

\begin{table}[ht]
\centering
\begin{tabular}[t]{|c|l|l|l|l|l|c|}
\hline
$\log_{10}(-q^2/M^2)$
&$M_{0;2}^{(1)}$&$M_{0;2}^{(2)}$&$M_{0;2}^{(3)}$
&$M_{0;2}^{(4)}$&$M_{0;2}^{(5)}$&$M_{0;2}^{(6)}$
\\[1.0ex]
\hline \hline

1 & 1.1434 & 0.3675 & 0.01199 & 0.0001163 &  7.7E-7   &  5.13E-9\\[1.0ex]
2 & 0.5766 & 0.5516 & 0.1065  & 0.006140  &  0.00015  &  2.31E-6\\[1.0ex]
3 & 0.3531 & 0.5360 & 0.2384  & 0.04244   &  0.0036   &  0.000177\\[1.0ex]
4 & 0.2305 & 0.4591 & 0.3200  & 0.1076    &  0.020    &  0.002358\\[1.0ex]
5 & 0.1592 & 0.3762 & 0.3456  & 0.1716    &  0.053    &  0.01093\\[1.0ex]
6 & 0.1154 & 0.3051 & 0.3370  & 0.2168    &  0.092    &  0.02811\\[1.0ex]
7 & 0.08700& 0.2484 & 0.3121  & 0.2410    &  0.129    &  0.05156\\[1.0ex]
8 & 0.06777& 0.2041 & 0.2818  & 0.2490    &  0.158    &  0.07691\\[1.0ex]    
9 & 0.05419& 0.1696 & 0.2513  & 0.2457    &  0.177    &  0.1006\\[1.0ex]
10& 0.04427& 0.1425 & 0.2229  & 0.2363    &  0.188    &  0.1196\\[1.0ex]
\hline
\end{tabular}
\caption{\footnotesize Values of moment $M_{0;2}^{(r)}$}
\label{mkrmomentsl0}
\end{table}

\begin{table}[ht]
\centering
\begin{tabular}[t]{|c|l|l|l|l|l|l|}
\hline
$\log_{10}(-q^2/M^2)$
&$M_{1;1}^{(1)}$&$M_{1;1}^{(2)}$&$M_{1;1}^{(3)}$
&$M_{1;1}^{(4)}$&$M_{1;1}^{(5)}$&$M_{1;1}^{(6)}$   
\\[1.0ex]
\hline \hline
1  & 0.5717  & 0.3822 & 0.02045 & 0.000075&$-1.7$E-6&$-5.0$E-8\\[1.0ex]
2  & 0.2883  & 0.5439 & 0.1055  & 0.00273 &$-5.6$E-5&$-3.3$E-6\\[1.0ex]   
3  & 0.1766  & 0.5565 & 0.2107  & 0.02029 &$-8.4$E-5&$-8.3$E-5\\[1.0ex]   
4  & 0.1153  & 0.5127 & 0.2879  & 0.05878 &$\phantom{-}0.00286$&$-0.00050$\\[1.0ex]   
5  & 0.0796  & 0.4545 & 0.3301  & 0.1072  &$\phantom{-}0.0139$&$-0.00069$\\[1.0ex]   
6  & 0.0577  & 0.3986 & 0.3451  & 0.1526  &$\phantom{-}0.0342$&$\phantom{-}0.00162$\\[1.0ex]   
7  & 0.04350 & 0.3501 & 0.3427  & 0.1885  &$\phantom{-}0.0601$&$\phantom{-}0.00848$\\[1.0ex]   
8  & 0.03389 & 0.3096 & 0.3307  & 0.2133  &$\phantom{-}0.0873$&$\phantom{-}0.02014$\\[1.0ex]   
9  & 0.02710 & 0.2760 & 0.3141  & 0.2282  &$\phantom{-}0.1123$&$\phantom{-}0.03531$\\[1.0ex]   
10 & 0.02214 & 0.2481 & 0.2959  & 0.2354  &$\phantom{-}0.1333$&$\phantom{-}0.05215$\\[1.0ex]   
\hline
\end{tabular}
\caption{\footnotesize Values of moment $M_{1;1}^{(r)}$}
\label{mkrmomentsl1}
\end{table}

\begin{table}[ht]
\centering
\begin{tabular}[t]{|c|c|c|c|c|}
\hline
$\log_{10}(-q^2/M^2)$
&$\sum_{k=1}^3M_{0;2}^{(2k-1)}$
&$\sum_{k=1}^3M_{0;2}^{(2k)}$
&$\sum_{k=1}^3M_{1;1}^{(2k-1)}$
&$\sum_{k=1}^3M_{1;1}^{(2k)}$
\\[1.0ex]
\hline \hline
1 &  1.155&  0.3676&  0.5921&  0.3823\\[1.0ex]
2 &  0.683&  0.5577&  0.3937&  0.5466\\[1.0ex]
3 &  0.595&  0.5786&  0.3872&  0.5767\\[1.0ex]
4 &  0.571&  0.5690&  0.4061&  0.5710\\[1.0ex]
5 &  0.558&  0.5587&  0.4236&  0.5610\\[1.0ex]
6 &  0.544&  0.5500&  0.4370&  0.5528\\[1.0ex]
7 &  0.528&  0.5410&  0.4463&  0.5471\\[1.0ex]
8 &  0.508&  0.5300&  0.4519&  0.5430\\[1.0ex]    
9 &  0.482&  0.5161&  0.4535&  0.5395\\[1.0ex]
10&  0.455&  0.4984&  0.4513&  0.5357\\[1.0ex]
\hline
\end{tabular}
\caption{\footnotesize Values of sums of moments $M_{0;2}^{(r)}$
and $M_{1;1}^{(r)}$, in the even and odd channels}
\label{mkrmomentstot}
\end{table}

\clearpage

\section{One--loop $2d$ integrals} 
\label{app1on}

We start with the 1--loop Euclidean integral with 2 internal 
scalar propagators with masses $m_1, m_2$:
\be
B(k;m_1,m_2)=\int_{-\infty}^{\infty}\frac{\rmd^2q}{(2\pi)^2}
\frac{1}{[(q+k)^2+m_1^2][q^2+m_2^2]}\,.
\end{equation}
The integral can be done analytically to obtain
\be
B(k;m_1,m_2)=
\frac{1}{2\pi \sqrt{\left(k^2+m_-^2\right)\left(k^2+m_+^2\right)}}
\ln\left\{\frac{\sqrt{k^2+m_+^2}+\sqrt{k^2+m_-^2}}
         {\sqrt{k^2+m_+^2}-\sqrt{k^2+m_-^2}}\right\}\,,
\end{equation}
where
\be
m_\pm=m_1\pm m_2\,.
\end{equation}

For the equal mass case $m_1=m_2=M$ we have
\ba
B(k)\equiv B(k;M,M)&=&\frac{1}{2\pi\sqrt{k^2(k^2+4M^2)}}
\ln\frac{\sqrt{k^2+4M^2}+\sqrt{k^2}}{\sqrt{k^2+4M^2}-\sqrt{k^2}}
\\
&=&b(\theta)=\frac{\theta}{4\pi M^2\sinh\theta}\,,\,\,\,\,
{\rm for}\,\,k^2=4M^2\sinh^2\frac{\theta}{2}\,.
\ea
Note $B(k)$ is analytic in $k^2$ with a cut from $-\infty$
to $-4M^2$. Also $B(k)\ne0$ for all $k^2$ and
\ba
B(k)&\sim&\frac{\ln k^2}{2\pi k^2}\,\,\,\,\,{\rm for}
\,\,\,k^2\to\infty\,,
\\
B(0)&=&\frac{1}{4\pi M^2}\,.
\ea
It can be represented by the dispersion relation
\ba
B(k)&=&\frac{1}{2\pi i}\int_{-\infty}^{-4M^2}\rmd z
\frac{B(z+i\epsilon)-B(z-i\epsilon)}{z-k^2}
\\
&=&\frac{1}{2\pi}\int_0^\infty \rmd\kappa
\frac{1}{k^2+4M^2\cosh^2\frac{\kappa}{2}}\,,
\ea
where we have substituted $z=-4M^2\cosh^2\frac{\kappa}{2}$
and noted $z\pm i\epsilon$ corresponds to setting $\theta=i\pi\pm\kappa$
with $\kappa>0$: 
\be
\frac{1}{2\pi i}\left[b(i\pi+\kappa)-b(i\pi-\kappa)\right]=
\frac{-1}{4\pi M^2\sinh\kappa}\,.
\end{equation}

The inverse of $B$ satisfies a once subtracted dispersion relation
\be
B(k)^{-1}=B(0)^{-1}
+\frac{k^2}{2\pi i}\int_{-\infty}^{-4M^2}\rmd z
\frac{B(z+i\epsilon)^{-1}-B(z-i\epsilon)^{-1}}{z(z-k^2)}\,.
\end{equation}
Noting
\be
\frac{1}{2\pi i}
\left[\frac{1}{b(i\pi+\kappa)}-\frac{1}{b(i\pi-\kappa)}\right]=
\frac{{4\pi M^2}\sinh\kappa}{\kappa^2+\pi^2}\,,
\end{equation}
we have
\be
B(k)^{-1}=4\pi M^2\left[1+2k^2\int_0^\infty \rmd\kappa
\frac{\sinh^2\frac{\kappa}{2}}{(\kappa^2+\pi^2)
\left(k^2+4M^2\cosh^2\frac{\kappa}{2}\right)}\right]\,.
\end{equation}

\subsection{General 1--loop integrals (``cutting rule")}

We consider an arbitrary 1--loop integral $(\sum_{i=1}^n k_i=0)$:
\be
I({\bf k})=\int\frac{\rmd^2 q}{(2\pi)^2}\prod_{i=1}^n
\left[(q+l_i)^2+m_i^2\right]^{-1}\,,
\end{equation}
where
\be
l_i=\sum_{j=1}^i k_j\,\,\,;\,\,(l_n=0)\,.
\end{equation}
The result is simply
\be
I({\bf k})=\sum_{i<j}\frac12\left(I^+_{ij}+I^-_{ij}\right)
B(l_{ij};m_i,m_j)\,,
\end{equation}
where
\be
I^\pm_{ij}=\prod^n_{r=1,r\ne i,j}
\left[(q+l_r)^2+m_r^2\right]^{-1}|_{q=q^\pm_{ij}}\,,
\end{equation}
and the momenta $q_{ij}^\pm$ are given by
\footnote{Note
$\left(q_{ij}^\pm+l_i\right)^2=-m_i^2$ and $q_{ij}^\pm=q_{ji}^\mp$\,.}
\ba
2q_{ij}^\pm&=&-(l_i+l_j)
-\frac{\left(m_i^2-m_j^2\right)}{l_{ij}^2}l_{ij}
\mp \frac{i}{l_{ij}^2}
\sqrt{s_{ij}^4+4m_j^2l_{ij}^2}\,\,\epsilon\cdot l_{ij}\,,
\\
l_{ij}&=&l_i-l_j\,,
\\
s_{ij}^2&=&l_{ij}^2+m_i^2-m_j^2\,.
\ea

\vfill
\eject



\eject


\begin{thebibliography}{99}

\bibitem{DIS1}
J.~Balog, P.~Weisz, Phys. Letts. B594 (2004) 141

\bibitem{Polyakov}
A.~M.~Polyakov, Phys. Lett. 72B (1977) 224

\bibitem{Luscher}
M.~L\"uscher,
Nucl. Phys. B135 (1978) 1

\bibitem{ZZ}
A.~B. and Al.~B.~Zamolodchikov,
Ann. Phys. 120 (1979) 253; Nucl. Phys. B133 (1978)  525

\bibitem{KaWe}
M.~Karowski, P.~Weisz,
Nucl. Phys. B139 (1978) 455

\bibitem{Smirnov}
F.~A.~Smirnov, Form factors in Completely Integrable Models
of Quantum Field Theory, World Scientific, 1992

\bibitem{Karowski}
M.~Karowski, {\it in} Field theoretical methods in particle physics,
1980, ed. W. R\"{u}hl, Pg. 307

\bibitem{SigmaI}
J.~Balog, M.~Niedermaier, F.~Niedermayer, 
A.~Patrascioiu, E.~Seiler, P.~Weisz, 
Phys. Rev. D60 (1999) 094508  

\bibitem{JanosMax}
J.~Balog, M.~Niedermaier, Phys. Rev. Lett. 78 (1997) 4151; 
Nucl. Phys. B500 (1997) 421 





\bibitem{BuVe}
D.~Buchholz, R.~Verch,
Rev. Math. Phys. 7 (1995) 1195; ibid 10 (1998) 775;
D.~Buchholz,
Nucl. Phys. B469 (1996) 333; also,
Talk at 12th International Congress of Math Phys (ICMP97),
Brisbane, Australia 13-19 July 1997, hep-th/9710094

\bibitem{CMP}
S.~Caracciolo, A.~Montanari, A.~Pelissetto, JHEP 0009 (2000) 045;
A.~Montanari, Ph.D.Thesis, hep-lat/0104005 

\bibitem{HMN} 
P.~Hasenfratz, M.~Maggiore, F.~Niedermayer,
Phys. Lett. B245 (1990) 522;
P.~Hasenfratz, F.~Niedermayer,
Phys. Lett. B245 (1990) 529

\bibitem{JanosPeter1}
J.~Balog, P.~Weisz,
Nucl. Phys. B668 (2003) 506

\bibitem{largenref}
M.~Moshe, J.~Zinn-Justin, 
Phys. Rept. 385 (2003) 69

\end{thebibliography}
\end{document}